\documentclass[twocolumn,showpacs,aps,prl]{revtex4-2}

\usepackage{mathrsfs}
\usepackage{amssymb, amsbsy, amsmath, latexsym, dsfont, array, layout, mathrsfs, color}

\usepackage{setspace}
\usepackage{braket}
\usepackage{physics}
 \usepackage{leftidx}

\usepackage{mathtools}

\usepackage{amsmath,amssymb}
\usepackage{bm}
\usepackage{graphicx}
\usepackage{braket,dsfont}
\usepackage{color}
\usepackage[10pt]{moresize}
\usepackage{mathrsfs}
\usepackage{multirow}
\usepackage{hyperref}
\usepackage{changes}
\usepackage{mathdots}
\let\oldcaption\caption
\renewcommand{\caption}{\sffamily \oldcaption}

\usepackage{amsmath,amsfonts,amssymb}
\usepackage{wrapfig}
\usepackage{graphicx}
\usepackage{bbm}
\usepackage[normalem]{ulem}
\usepackage{color}

\begin{document}

\title{Preparing Remote States for Genuine Quantum Networks}

\author{Shih-Hsuan Chen$^{1,2}$}
\author{Chan Hsu$^{1,2}$}
\author{Yu-Chien Kao$^{1,2}$}
\author{Bing-Yuan Lee$^{1,2}$}
\author{Yuan-Sung Liu$^{1,2}$}
\author{Yueh-Nan Chen$^{2,3}$}
\author{Che-Ming Li$^{1,2,4,}$}
\email{cmli@mail.ncku.edu.tw}

\affiliation{$^{1}$Department of Engineering Science, National Cheng Kung University, Tainan 70101, Taiwan}
\affiliation{$^{2}$Center for Quantum Frontiers of Research \& Technology, National Cheng Kung University, Tainan 70101, Taiwan}
\affiliation{$^{3}$Department of Physics, National Cheng Kung University, Tainan 70101, Taiwan}
\affiliation{$^{4}$Center for Quantum Science and Technology, Hsinchu 30013, Taiwan}

\begin{abstract}
Quantum networks typically comprise quantum channels, repeaters, and end nodes. Remote state preparation (RSP) allows one end node to prepare the states of the other end nodes remotely. While quantum discord has recently been recognized as necessary for RSP, it does not guarantee the practical implementation of RSP in quantum networks surpasses any classical method. Herein, we theoretically introduce and experimentally study a quantum resource that we call the RSP capability. This resource validates all the static and dynamic elements required to enable genuine quantum networks where the RSP's implementation can outperform any classical emulation of entanglement- and qubit-unitaries-free strategies, including the static resources of Einstein-Podolsky-Rosen pairs and the dynamic resources of quantum channels and repeaters. Our experiment measures the RSP capability to demonstrate the transition between classical and nonclassical RSP depending on the photon-pair qualities. It shows that quantum discord does not confirm a nonclassical RSP, but the RSP capability does. These results help reveal the quantum advantages that emerge when networking RSP is in play.
\end{abstract}

\maketitle

\section*{Introduction}


The remote state preparation (RSP) \cite{Pati00,Bennett01} process confers the advantage of remotely manipulating the qubit of a receiver node (Bob) by a sender node (Alice). In particular, if Bob cannot implement a target operator locally, Alice can help. Moreover, compared to teleportation \cite{Bennett93}, which requires Bell-state measurements \cite{Pirandola15} to transmit an unknown qubit, the RSP protocol requires only local measures on Alice’s particle. These two features make RSP uniquely appealing and well-suited for realizing quantum networking sources and elements, from deterministically creating single-photon states \cite{Jeffrey04} to preparing single-photon hybrid entanglement \cite{Barreiro10} and initializing atomic quantum memory for quantum communication \cite{Bao12}. For the role of RSP in preparing a quantum memory state, RSP also becomes essential in quantum-memory-related applications such as memory-assisted measurement-device-independent quantum key distribution \cite{Panayi14} and space-borne quantum memories for global quantum networking \cite{Gundo21,Walln22,Gundo23}, being helpful for photon-mediated quantum networks consisting of quantum channels, repeaters, and end nodes.

Moreover, RSP is a faithful quantum channel that is essential to assist in performing client-server blind quantum computation \cite{Gustiani2021}. Similar principles to those underlying RSP also play a role in a critical range of measurement-based quantum information processes, such as one-way quantum computing \cite{raussendorf2001,briegel2009,you2007,tanamoto2009,wang2010} and entanglement-enabled networks \cite{Chou18,Pirker18,Wehner18} for global networking and pathfinder missions \cite{Gundo21_2,Colquhoun22,Islam22,Sidhu21}. Similar to the RSP, one-way quantum computing is implemented by performing measurements on partial qubits of a highly entangled cluster state to output the rest of the qubits of the cluster state as computation results. Entanglement-enabled networks can realize such networking tasks and the related extensions \cite{Barreiro10} by using more functionalities for handling superposed tasks and superposed addressing \cite{Miguel-Ramiro21}. Understanding RSP promotes a more profound understanding of these general measurement-based quantum-information processing.

As summarized above, RSP is vital for quantum networks. Since the experimental realization of RSP always involves noise and imperfections, identifying whether a practical implementation of RSP in quantum networks surpasses any classical method is of great interest and essential when examining the related quantum hardware and resources for experimentally realizing genuine quantum networks. We expect an RSP-based genuine quantum network in experimental realization to possess the quantum functionality of preparing and transmitting quantum information that goes beyond the predictions of classical theory. The terminology of genuine quantum networks distinguishes the quantum networks derived from other quantum characteristics that cannot support the nonclassical RSP. The existing verification methods \cite{Dakic12,Dakic10,Luo10,Killoran10} based on quantum discord \cite{Ollivier01,Zurek00,Henderson01} or RSP benchmarks \cite{Killoran10} cannot determine whether a practical implementation of RSP is truly quantum and distinct from the results of classical theories. As shown below, quantum discord does not guarantee the nonclassical RSP. These verifications rely on
assumptions deduced from quantum theory. They do not generally hold for practical implementations where classical physics can describe experimental imperfections. Therefore, one cannot use these tools to confirm the implementation of nonclassical state preparation and transmission for genuine quantum networks.

 Confirming a genuinely quantum implementation hinges on how generic classical processes can perform RSP and RSP's applications. This work develops a general classical RSP model without Einstein-Podolsky-Rosen (EPR) entangled pairs and qubit unitaries. We derive a quantum resource that we call the RSP capability, which is required for nonclassical state preparation in general quantum networks. This resource is distinct from the static resources of quantum states, such as quantum correlations. Experimentally, we use our classical RSP model to measure the RSP capability for practical RSP verifications. Our demonstration shows that while polarization-correlated photon pairs possess quantum discord and standard Einstein-Podolsky-Rosen (EPR) steering \cite{Wiseman07} for implementing RSP, the final state preparation processes do not outperform the best classical emulation of RSP. We also reveal that, even though photon pairs have quantum discord, the corresponding resulting state preparation does not have the RSP capability, and the classical model can still describe it.

\section*{Results}

\subsection*{Quantum operation of remote state preparation (RSP)}

\begin{figure}[t]
\includegraphics[width=7.9cm]{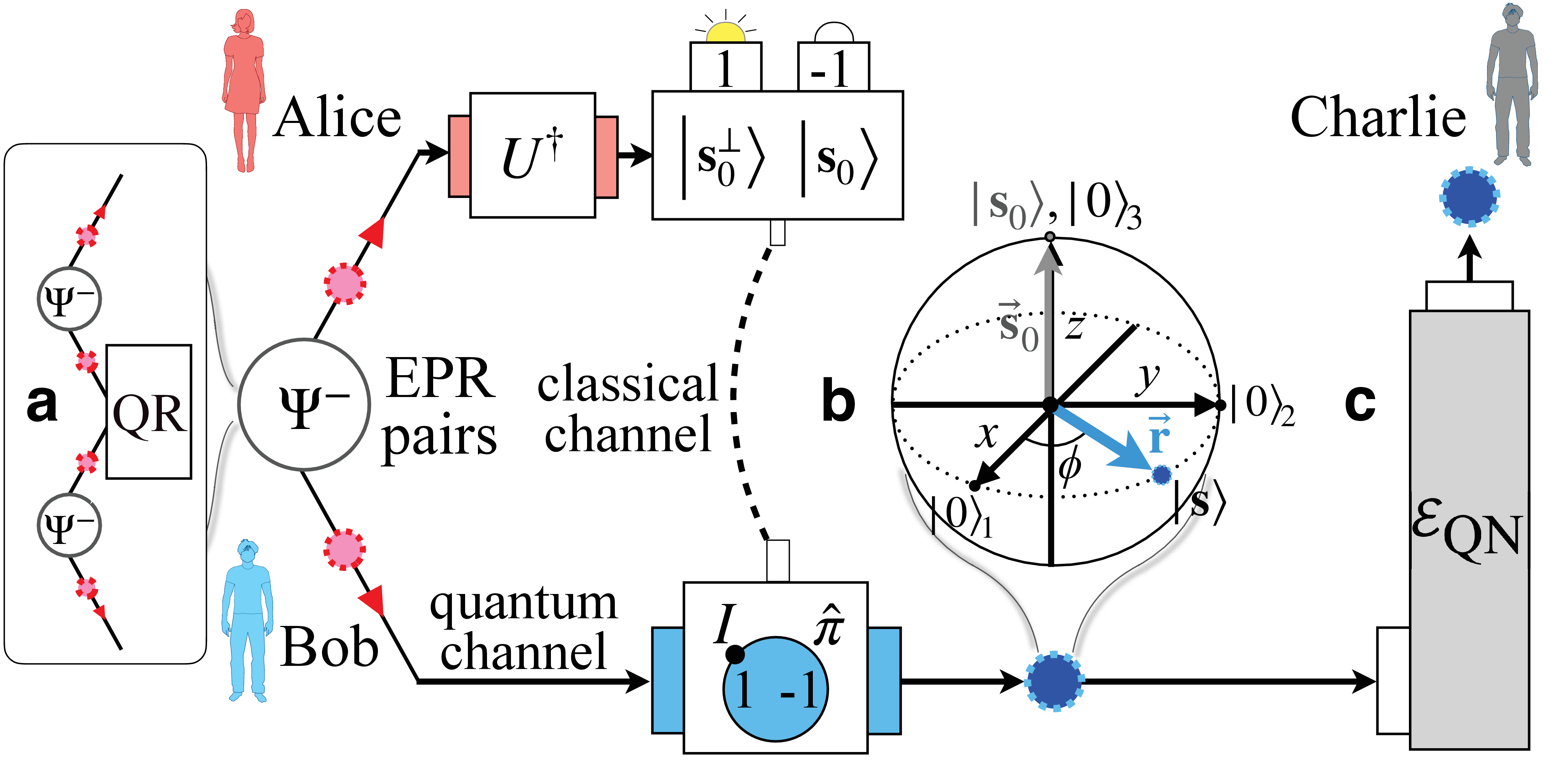}
\caption{Remote state preparation (RSP) in a quantum network. \textbf{a} Two distant nodes, Alice and Bob, first use quantum repeaters (QRs) to share Einstein-Podolsky-Rosen (EPR) entangled pairs for long-distance communication. \textbf{b} To prepare Bob’s remote state: $\ket{\bold{s}}=U\ket{\bold{s}_0}$, as shown on the equatorial of the Bloch sphere \cite{Nielsen00}, Alice performs the operation $U^{\dag}$ and then measures her qubit in the basis $\{\ket{\bold{s}_{0}},\ket{\bold{s}_{0}^{\bot}}\}$. Here, the states $\ket{0}_1$, $\ket{0}_2$, and $\ket{0}_3$ on the Bloch sphere denote the eigenvectors of the Pauli-$X$, Pauli-$Y$, and Pauli-$Z$ observables, respectively, with the eigenvalue $1$. If Alice obtains the measurement outcome $\ket{\bold{s}_{0}^{\bot}}$ ($\ket{\bold{s}_{0}}$), she sends Bob the message $1$ ($-1$) via a classical channel to show that his correction unitary operator, $\hat{\pi}$, is unnecessary (necessary). Here, the identity operator $I$ represents doing nothing. The RSP process can be considered a quantum operation, Eq.~(\ref{RSPQO}), transforming the initial state $\ket{\bold{s}_0}$ to the final state $\ket{\bold{s}}$, corresponding to the rotation from the Bloch vector $\vec{\bold{s}}_0$ to the final one, $\vec{\bold{r}}$. \textbf{c} With an extended quantum channel or quantum network ($\mathcal{E}_{{\rm QN}}$) consisting of teleportation and (or) QR, Bob can send the prepared state $\ket{\bold{s}}$ to a third distant end node, Charlie.}
\label{RSP}
\end{figure}

The RSP protocol \cite{Pati00,Bennett01} uses static and dynamic elements in a quantum network. Alice and Bob initially share the static resources of rotationally symmetric entangled Einstein-Podolsky-Rosen (EPR) pairs using quantum repeaters (QRs) (Fig.~\ref{RSP}\textbf{a}). The EPR pairs' state vectors for rotational symmetry can be in the following form\cite{Cabello02}:
\begin{equation}
\ket{\Psi^{-}}=\frac{1}{\sqrt{2}}(U\ket{\bold{s}_{0}}\otimes U\ket{\bold{s}_{0}^{\bot}}-U\ket{\bold{s}_{0}^{\bot}}\otimes U\ket{\bold{s}_{0}}),\nonumber
\end{equation}
where $U$ is an arbitrary single-qubit unitary operator, and $\ket{\bold{s}_{0}}$ and $\ket{\bold{s}_{0}^{\bot}}$ form an orthonormal basis. In the RSP protocol, suppose Alice plans to prepare a state on the equator of the Bloch sphere for Bob, say $\ket{\bold{s}}=(\ket{\bold{s}_{0}}+e^{i\phi}\ket{\bold{s}_{0}^{\bot}})/\sqrt{2}$ (Fig.~\ref{RSP}\textbf{b}). She follows the following two steps to achieve the aim.

First, she measures her qubit of $\ket{\Psi^{-}}$ in the basis $\{U\ket{\bold{s}_{0}},U\ket{\bold{s}_{0}^{\bot}}\}$, where $U$ is a rotation operator that can perform the state transformation: $\ket{\bold{s}}=U\ket{\bold{s}_{0}}$. To perform this measurement, she first applies the operation $U^\dag$ on her qubit and then measures it in the basis $\{\ket{\bold{s}_{0}},\ket{\bold{s}_{0}^{\bot}}\}$. See Fig.~\ref{RSP}\textbf{b}. Implementing the operation $U^\dag$ can be considered a dynamic resource Alice requires in the RSP protocol. She has equal probabilities of observing two orthogonal states, either $\ket{\bold{s}_{0}}$ or $\ket{\bold{s}_{0}^{\bot}}$. As Alice observes $\ket{\bold{s}_{0}}$, Bob's qubit is in the state $U\ket{\bold{s}_{0}^{\bot}}$. Whereas when she observes $\ket{\bold{s}_{0}^{\bot}}$, Bob's state is in the state $U\ket{\bold{s}_{0}}$. Second, depending on the measurement outcome, Alice sends Bob a message to tell him whether he needs to perform a $\pi$ rotation operator $\hat{\pi}$ about the $z$-axis of the Bloch sphere on his qubit for finalizing state preparation. Here, classical communication and Bob's correction are also dynamic resources Alice and Bob require. Alice informs Bob that such a correction is unnecessary if Alice observes $\ket{\bold{s}_{0}^{\bot}}$ (Fig.~\ref{RSP}\textbf{b}.). However, when Alice observes $\ket{\bold{s}_{0}}$, she informs Bob that the correction is necessary, that is $\hat{\pi}U\ket{\bold{s}_{0}^{\bot}}=\ket{\bold{s}}$. According to the above two steps, the final state Bob obtains using the RSP protocol is:
\begin{equation}
\frac{1}{2}U\ketbra{\bold{s}_{0}}U^\dag+\frac{1}{2}\hat{\pi}U\ketbra{\bold{s}_{0}^{\bot}}U^\dag\hat{\pi}^\dag=U\ketbra{\bold{s}_{0}}U^\dag.\label{ideal}
\end{equation}

It is worth noting that the above RSP protocol can be modified to a version without using Bob's correction operations. Alice's specific measurement outcome $\ket{\bold{s}_{0}^{\bot}}$ is used as a heralding signal, showing that Bob's state has been successfully prepared. When Alice observes the state $\ket{\bold{s}_{0}}$, they skip the present state preparation round. The original deterministic RSP becomes a heralded RSP, which is especially useful when it has difficulty realizing Bob's corrections, such as in the quantum memory system~\cite{Bao12,Panayi14}. Here, Alice and Bob still need classical communication to confirm the observation of a heralding signal. The final state of Bob's qubit is
\begin{equation}
\frac{\tr_{A}(\ketbra{\Psi^{-}}\hat{M}_A)}{\tr(\ketbra{\Psi^{-}}\hat{M}_A)}=U\ketbra{\bold{s}_{0}}U^\dag,\label{hed}
\end{equation}
where $\hat{M}_A=U\ketbra{\bold{s}_{0}^{\bot}}U^\dag$ represents the measurement operator for Alice's conditional measurement, and $\tr(\ketbra{\Psi^{-}}\hat{M}_A)$ is the heralding (or success) probability. Consider the RSP used in an atomic-ensemble quantum memory initialization as a concrete example. Applying a write pulse can create a pair of entanglement between the spin wave vector (single collective atomic excitation) and the momentum of the write-out photon through Raman scattering \cite{Chen07}. After converting the momentum degree of the write-out photon into the polarization degree, one can create the entanglement between the spin wave state of the ensemble and the polarization of the write-out photon. As experimentally demonstrated in Ref.~\cite{Bao12}, performing the measurement of $\hat{M}_A$ on the photon in the created atom–photon entangled state can prepare the atomic ensemble memory as a specific state of $U\ket{\bold{s}_{0}}$ for later applications, such as teleportation \cite{Bao12}.

Thus, according to Eqs.~(\ref{ideal}) and (\ref{hed}) the deterministic RSP and the heralded version can be represented as a quantum operation of unitary transformation for all considered pure input states $\rho_{\bold{s}_{0}}=\ketbra{\bold{s}_{0}}$ of RSP:
\begin{equation}
\mathcal{E}_{\text{RSP}}(\rho_{\bold{s}_{0}})=U\rho_{\bold{s}_{0}}U^\dag=\rho_{\bold{s}},\label{RSPQO}
\end{equation}
where $\rho_{\bold{s}}=U\ketbra{\bold{s}_0}U^{\dag}=\ketbra{\bold{s}}$ are considered the output states of $\mathcal{E}_{\text{RSP}}$. As discussed above, Eq.~(\ref{RSPQO}) results from all the static and dynamic resources involved in preparing remote states. While the RSP process has a quantum operation interpretation comparable to quantum teleportation, the physical resources and elements required for these processes differ. We can see such a difference from deriving the teleportation’s quantum operation. Compared to Eq.~(\ref{RSPQO}) for RSP, we have four different Bell-state measurement events and the corresponding four different kinds of strategies in teleportation, with or without conditional corrections, to obtain the following input-output relation of teleportation’s quantum operation \cite{Pirandola15,Bao12}: $\mathcal{E}_{\text{Tel}}(\rho_{\bold{s}_{0}})=(I\ketbra{\bold{s}_{0}} I^\dag+\sigma_1X\ketbra{\bold{s}_{0}} X^\dag\sigma^\dag_1+\sigma_2Y\ketbra{\bold{s}_{0}} Y^\dag\sigma^\dag_2+\sigma_3Z\ketbra{\bold{s}_{0}} Z^\dag\sigma^\dag_3)/4=I\rho_{\bold{s}_{0}}I^\dag=\rho_{\bold{s}_{0}}$, where $\rho_{\bold{s}_{0}}$ is the teleported state, $I$ is the idenity operator, and $\sigma_1=X$, $\sigma_2=Y$, and $\sigma_3=Z$ are the Pauli-$X$, Pauli-$Y$, and Pauli-$Z$ matrices, respectively. Although the resulting forms of $\mathcal{E}_{\text{RSP}}$ and $\mathcal{E}_{\text{Tel}}$ involve only single-qubit unitaries, $\mathcal{E}_{\text{Tel}}$ requires much heavier dynamic resources.

\subsection*{RSP process in quantum networks}
Equation~(\ref{RSPQO}) implies that the distant Alice and Bob cooperate to realize the RSP quantum operation, $\mathcal{E}_{\text{RSP}}$, which outputs a qubit with a well-defined state deterministically or heraldedly at Bob's side. RSP is highly adaptive in photon-mediated quantum networks~\cite{Wehner18} for preparing and transmitting single photons to another end node. Therefore, we use the following concatenated quantum operation to describe the RSP-assisted state preparation and transmission in a quantum network:
\begin{equation}
\mathcal{E}_{\text{QN}}\circ\mathcal{E}_{\text{RSP}}(\rho_{\bold{s}_{0}})=\rho_{\bold{s}},\label{RSPQN}
\end{equation}
where $\mathcal{E}_{\text{QN}}$ denotes the state-preservation quantum operation of the photon-mediated quantum channel or network, and the symbol $\circ$ describes that $\mathcal{E}_{\text{QN}}$ concatenates $\mathcal{E}_{\text{RSP}}$. In our quantum networking scheme, a quantum channel $\mathcal{E}_{\text{QN}}$ can generally consist of subchannels such as teleportation and (or) quantum repeater. Then, the state prepared at Bob’s hand with Alice’s help can be transmitted by utilizing quantum communication protocols, such as teleportation, to send from Bob to a third party. See Fig.~\ref{RSP}\textbf{c} for an illustration of $\mathcal{E}_{\text{QN}}$, where Charlie is the end node.

As experimentally demonstrated in Ref.~\cite{Bao12}, teleportation is the target extended quantum channel, $\mathcal{E}_{\text{QN}}=\mathcal{E}_{\text{Tel}}$. It transmits the state of a quantum memory prepared by the RSP from one node (Bob) to another atomic quantum memory node (Charlie). Equation~(\ref{RSPQN}) concatenates the prepared state $\mathcal{E}_{\text{RSP}}(\rho_{\bold{s}_{0}})=\rho_{\bold{s}}$ at Bob's side and the resulting transmitted state $\mathcal{E}_{\text{Tel}}(\rho_{\bold{s}})=\rho_{\bold{s}}$ at Charlie's side in this example's ideal case, i.e., $\mathcal{E}_{\text{Tel}}\circ\mathcal{E}_{\text{RSP}}(\rho_{\bold{s}_{0}})=\mathcal{E}_{\text{Tel}}(\rho_{\bold{s}})=\rho_{\bold{s}}$.

To describe the RSP process $\mathcal{E}_{\text{RSP}}$, the concatenated process $\mathcal{E}_{\text{QN}}\circ\mathcal{E}_{\text{RSP}}$, and their corresponding experimental realizations, denoted as $\mathcal{E}$, in a quantatively precise manner using experimentally obtainable data, we utilize the quantum operations formalism \cite{Nielsen00} to specify the requirements and characteristics of these processes in matrix forms, called the process matrices. According to the process tomography of quantum operations formalism, an experimental process's (un-normalized) process matrix must be positive Hermitian in the following form \cite{Nielsen00}:

\begin{widetext}
\begin{equation}
\tilde{\chi}_{\mathcal{E}}\!=\!\Lambda
\left[
\begin{matrix}
\mathcal{E}(\ket{0}_{\!33}\!\!\bra{0}) &\!\!\mathcal{E}(\!\ket{0}_{\!11}\!\!\bra{0}\!)\!+\!i\mathcal{E}(\!\ket{0}_{\!22}\!\!\bra{0}\!)\!-\!\frac{1+i}{2}[\mathcal{E}(\!\ket{0}_{\!33}\!\!\bra{0}\!)\!+\!\mathcal{E}(\!\ket{1}_{\!33}\!\!\bra{1}\!)]\! \\
\!\mathcal{E}(\!\ket{0}_{\!11}\!\!\bra{0}\!)\!-\!i\mathcal{E}(\!\ket{0}_{\!22}\!\!\bra{0}\!)\!-\!\frac{1-i}{2}[\mathcal{E}(\!\ket{0}_{\!33}\!\!\bra{0}\!)\!+\!\mathcal{E}(\!\ket{1}_{\!33}\!\!\bra{1}\!)] & \mathcal{E}(\ket{1}_{\!33}\!\!\bra{1})\!\\
\end{matrix}
\right]\!\Lambda,\label{pt}
\end{equation}
\end{widetext}
where 
\begin{equation}
\Lambda=\frac{1}{2}
\left[
\begin{matrix}
I & X \\
X & -I\\
\end{matrix}
\right],\nonumber
\end{equation}
and $\ket{n}_{m}$ are the eigenvectors of $\sigma_{m}$ with the eigenvalues $v_{nm}=(-1)^{n}$ for $n=0,1$ and $m=1,2,3$, $\ket{n}_1=(\ket{0}_3+v_{n1}\ket{1}_3)/\sqrt{2}$, and $\ket{n}_2=(\ket{0}_3+v_{n2}i\ket{1}_3)/\sqrt{2}$. In the Methods section, we have detailed the main steps for implementing the RSP protocol and how to get the experimental process matrix (\ref{pt}). 

Take the ideal RSP, for example. We have $\mathcal{E}=\mathcal{E}_{\text{RSP}}$ and use Eq.~(\ref{RSPQO}) to describe the output states $\mathcal{E}_{\text{RSP}}(\ket{n}_{mm}\!\bra{n})$. The process matrix $\tilde{\chi}_{\mathcal{E}}$ (\ref{pt}) becomes 
\begin{equation}
\tilde{\chi}_{\mathcal{E}}\!=\!
\Lambda\left[
\begin{matrix}
U\ket{0}_{\!33}\!\!\bra{0}U^{\dag} & U\ket{0}_{\!33}\!\!\bra{1}U^{\dag}\! \\
U\ket{1}_{\!33}\!\!\bra{0}U^{\dag} & U\ket{1}_{\!33}\!\!\bra{1}U^{\dag}\!\\
\end{matrix}
\right]\!\Lambda.\label{ptrsp}
\end{equation}
Since the concatenated quantum operation $\mathcal{E}_{\text{QN}}\circ\mathcal{E}_{\text{RSP}}$ has the same input-output relation as the $\mathcal{E}_{\text{RSP}}$ [Eq.~(\ref{RSPQN})], its process matrix is in the same form (\ref{ptrsp}). Moreover, when $U=I$, the quantum operation $\mathcal{E}_{\text{RSP}}$ has the same input-output relation as the teleportation $\mathcal{E}_{\text{Tel}}$: $\mathcal{E}_{\text{RSP}}(\rho_{\bold{s}_{0}})=\mathcal{E}_{\text{Tel}}(\rho_{\bold{s}_{0}})=\rho_{\bold{s}_{0}}$, and their process matrices become the same, $\tilde{\chi}_{\mathcal{E}_{\text{Tel}}}=\tilde{\chi}_{\mathcal{E}_{\text{RSP}}}$.

The $\tilde{\chi}_{\mathcal{E}}$ is experimentally measurable when the states $\mathcal{E}(\ket{n}_{mm}\!\bra{n})$ can be determined experimentally by quantum state tomography \cite{Nielsen00}. The process matrix contains all the process details regarding $\mathcal{E}(\ket{n}_{mm}\!\bra{n})$ required to describe output states. When $\tilde{\chi}_{\mathcal{E}}$ is known, the output states $\mathcal{E}(\rho_{\bold{s}_{0}})$ are also determined. See Supplementary Note 1 for the relation between $\tilde{\chi}_{\mathcal{E}}$ and $\mathcal{E}(\rho_{\bold{s}_{0}})$. The following simple analysis shows this. We first decompose a given input state $\rho_{\bold{s}_{0}}$ into a sum of the Pauli observables, called the state tomography: $\rho_{\bold{s}_{0}}=(I+\vec{\bold{s}}_{0}\cdot \vec{\sigma})/2$, where $\vec{\bold{s}}_{0}=(s_{01},s_{02},s_{03})$ with $s_{0m}=\tr(\sigma_m\rho_{\bold{s}_{0}})$ is the Bloch vector of $\rho_{\bold{s}_{0}}$ and $\vec{\sigma}=(\sigma_{1},\sigma_{2},\sigma_{3})$. The output state of an experimental process $\mathcal{E}$ is
\begin{equation}
\mathcal{E}(\rho_{\bold{s}_{0}})=\frac{1}{2}[\mathcal{E}(I)+\sum_{m=1}^{3}s_{0m}\mathcal{E}(\sigma_{m})],\label{st}
\end{equation}
where $\mathcal{E}(I)=\sum_{n=0}^{1}\mathcal{E}(\ket{n}_{mm}\!\!\bra{n})$ for arbitrary $m$ and $\mathcal{E}(\sigma_{m})=\sum_{n=0}^{1}v_{nm}\mathcal{E}(\ket{n}_{mm}\!\!\bra{n})$. Thus, $\tilde{\chi}_{\mathcal{E}}$ can provide all the knowledge about $\mathcal{E}(\ket{n}_{mm}\!\!\bra{n})$ required to construct the output state $\mathcal{E}(\rho_{\bold{s}_{0}})$ according to the above state tomography.

It is worth emphasizing that the static and dynamic resources involved in the networking RSP processes contribute to the resulting quantum operations. The process matrix $\tilde{\chi}_{\mathcal{E}_{\text{RSP}}}$ and related extended ones can manifest the effects of the whole underlying hardware and resources. In the following, we will introduce our classical networking RSP model and show how the results of undesired effects, including experimental imperfections or noise, can be characterized by comparing them with classical RSP process matrices. This theory enables the best classical simulations to satisfy the requirements of the quantum operations formalism.

\subsection*{Classical networking RSP}
In our classical networking RSP model, Alice and Bob perform the process without EPR pairs and any unitary transformations, including Alice's $U^{\dag}$ and Bob’s correction operations. 
They proceed only according to the following classical assumption about their shared states and the related classical communication. We then derive a classical networking RSP process matrix from the introduced classical RSP process.

\subsubsection*{Assumption of classical shared pairs and communication}
For each round of process concerning RSP~(\ref{RSPQO}) or overall networking transmission~(\ref{RSPQN}),
Alice and Bob receive a pair of $(v_\lambda,\rho_{\lambda})$ from a classical source. Alice has an instruction of pre-existing classical data (state), $v_\lambda$, and Bob or Charlie has a qubit of the density operator, $\rho_{\lambda}$, with a probability distribution $p(\lambda)$ for the shared pairs of $(v_\lambda,\rho_{\lambda})$. Here, $v_{\lambda}=(v_{n1},v_{n2},v_{n3})$ is a pre-existing measurement outcome set consisting of the three measurement results independent of observation, $v_{n1}$, $v_{n2}$, and $v_{n3}$, for $v_{nm}\in\{+1,-1\}$. We use the classical assumption of realism \cite{Brunner14} to describe that the three physical properties corresponding to the three observables $\sigma_m$ have three definite values $v_{nm}$ independent of observation. Eight possible pre-existing measurement outcome sets exist for these three physical properties of interest:
\begin{eqnarray}
v_1=(+1,+1,+1)&,&\ v_2=(+1,+1,-1),\ v_3=(+1,-1,+1),\ \nonumber\\ 
v_4=(+1,-1,-1)&,&\ v_5=(-1,+1,+1),\ v_6=(-1,+1,-1),\ \nonumber\\ 
v_7=(-1,-1,+1)&,&\ v_8=(-1,-1,-1).\nonumber
\end{eqnarray}
Therefore, eight possible Bob's qubit states $\rho_\lambda$ correspond to Alice's eight instructions of pre-existing classical data $v_\lambda$. Moreover, our model allows imperfect preparation of the shared pairs where Alice and Bob have a probability of unsuccessfully receiving the shared pairs $(v_\lambda,\rho_{\lambda})$ from the classical source, where the sum of the probability distribution is unnecessarily unit: $\sum_\lambda p(\lambda)\leq1$. Such a case describes the scenarios of Alice, Bob, or both who do not receive their instruction or qubit due to possible experimental imperfections causing information loss with a probability: $1-\sum_\lambda p(\lambda)$.

In each round of the classical RSP, when Alice receives an instruction $v_{\lambda}$ and Bob gets a qubit of $\rho_\lambda$ from the classical source of the shared pair $(v_\lambda,\rho_{\lambda})$, Alice will read her instruction $v_{\lambda}$ and randomly pick one from the three pre-existing measurement outcomes $v_{nm}$ to send to Bob through a one-bit classical communication channel. The average state Bob or Charlie receives conditioned on Alice's outcome $v_{nm}$ is
\begin{equation}
\tilde{\rho}_{c|v_{nm}}=\sum_{\lambda}p(\lambda|v_{nm})\rho_{\lambda} \ \ \ \forall \ n,m.\label{ec}
\end{equation}
Since the relation $p(\lambda|v_{nm})p(v_{nm})=p(v_{nm}|\lambda)p(\lambda)$ holds, we can rephrase the above equation in the following forms to emphasize the role of the probability distribution $p(\lambda)$:
\begin{eqnarray}
&&\tilde{\rho}_{c|v_{01}}=\sum_{\lambda=1,2,3,4}2p(\lambda)\rho_{\lambda},\ \tilde{\rho}_{c|v_{11}}=\sum_{\lambda=5,6,7,8}2p(\lambda)\rho_{\lambda},\ \nonumber\\
&&\tilde{\rho}_{c|v_{02}}=\sum_{\lambda=1,2,5,6}2p(\lambda)\rho_{\lambda},\ \tilde{\rho}_{c|v_{12}}=\sum_{\lambda=3,4,7,8}2p(\lambda)\rho_{\lambda},\ \label{ecall}\\
&&\tilde{\rho}_{c|v_{03}}=\sum_{\lambda=1,3,5,7}2p(\lambda)\rho_{\lambda},\ \tilde{\rho}_{c|v_{13}}=\sum_{\lambda=2,4,6,8}2p(\lambda)\rho_{\lambda},\nonumber
\end{eqnarray}
where we have assumed that each state $v_{nm}$ has an equal probability of $p(v_{nm})=1/2$ as the input state for determining the process matrix (\ref{pt}). Here, we also have used the fact that $p(v_{nm}|\lambda)=0$ or $1$, which depends on whether we can find the result $v_{nm}$ from the pre-existing measurement outcome set $v_{\lambda}=(v_{n1},v_{n2},v_{n3})$. Taking $p(v_{01}|\lambda)$ in the state $\tilde{\rho}_{c|v_{01}}$ in Eq.~(\ref{ecall}) for example, where $v_{01}=+1$, we have $p(v_{01}|\lambda)=1$ for $\lambda=1,2,3,4$ and $p(v_{01}|\lambda)=0$ for $\lambda=5,6,7,8$.

\subsubsection*{Classical networking RSP process matrix}
We use the following comparisons to construct the classical networking RSP process matrix. Alice's outcome $v_{nm}$ corresponds to the input state of a quantum operation, $\ket{n}_{mm}\!\bra{n}$, and Bob's or Charlie's qubit's prepared state, $\tilde{\rho}_{c|v_{nm}}$, corresponds to the output state of the quantum operation, $\mathcal{E}(\ket{n}_{mm}\!\bra{n})$. By substituting the states $\tilde{\rho}_{c|v_{nm}}$ (\ref{ec}) prepared through the classical procedure introduced above into the process matrix~(\ref{pt}), we arrive at the following classical networking RSP process matrix:
\begin{widetext}
\begin{equation}
\tilde{\chi}_{\mathcal{E}_c}\!=\!
\Lambda\left[
\begin{matrix}
\tilde{\rho}_{c|v_{03}} &\tilde{\rho}_{c|v_{01}}+i\tilde{\rho}_{c|v_{02}}-\frac{1+i}{2}(\tilde{\rho}_{c|v_{03}}+\tilde{\rho}_{c|v_{13}}) \\
\!\tilde{\rho}_{c|v_{01}}-i\tilde{\rho}_{c|v_{02}}-\frac{1-i}{2}(\tilde{\rho}_{c|v_{03}}+\tilde{\rho}_{c|v_{13}}) & \tilde{\rho}_{c|v_{13}}\!\\
\end{matrix}
\right]\Lambda,\label{pct}
\end{equation}
\end{widetext}
where $\text{tr}(\tilde{\chi}_{\mathcal{E}_c})=\sum_{\lambda}p(\lambda)\text{tr}(\rho_{\lambda})=\sum_{\lambda}p(\lambda)\leq1$. It is worth noting that $\text{tr}(\tilde{\chi}_{\mathcal{E}_c})<1$ corresponds to non-trace-preserving processes \cite{Nielsen00}. The classical networking RSP process matrix (\ref{pct}) defines a quantum operation of the classical networking RSP, $\mathcal{E}_c$. Explicitly, according to Eq.~(\ref{st}), $\mathcal{E}_{c}$ transforms an input state $\rho_{\bold{s}_{0}}$ into $\mathcal{E}_{c}(\rho_{\bold{s}_{0}})=\tilde{\rho}_{c|\bold{s}_{0}}$, where
\begin{equation}
\tilde{\rho}_{{c}|\bold{s}_{0}}=\frac{1}{2}(\sum_{n=0}^{1}\tilde{\rho}_{c|v_{nm}}+\sum_{m=1}^{3}\sum_{n=0}^{1}s_{0m}v_{nm}\tilde{\rho}_{c|v_{nm}}).\label{rrc}
\end{equation}
Equations~(\ref{ec},\ref{rrc}) imply that if a general quantum network contains a classical subnetwork, the whole network can also be considered a classical network. That is,
\begin{equation}
\mathcal{E}_{\text{QN}}\circ\mathcal{E}_{c}(\rho_{\bold{s}_{0}})=\tilde{\rho}_{{c}|\bold{s}_{0}},\text{or } \mathcal{E}_{c}\circ\mathcal{E}_{\text{RSP}}(\rho_{\bold{s}_{0}})=\tilde{\rho}_{{c}|\bold{s}},\label{RSPQNn}
\end{equation}
where $\mathcal{E}_{\text{QN}}=\mathcal{E}_{\text{QN}N}\circ...\circ\mathcal{E}_{\text{QN1}}$ consists of $N$ subnetworks. We obtain the former equality by: $\mathcal{E}_{\text{QN}}\circ\mathcal{E}_{c}(\ket{n}_{\!mm}\!\!\bra{n})=\mathcal{E}_{c}(\ket{n}_{\!mm}\!\!\bra{n})=\tilde{\rho}_{c|v_{nm}}$ via Eq.~(\ref{ec}) and the definition of $\tilde{\rho}_{{c}|\bold{s}_{0}}$ (\ref{rrc}). We get the latter by: $ \mathcal{E}_{\text{RSP}}(\rho_{\bold{s}_{0}})=\rho_{\bold{s}}$ and $ \mathcal{E}_{c}(\rho_{\bold{s}})=\tilde{\rho}_{{c}|\bold{s}}$. Without losing any generality, we first assume that RSP is ideal such that $\mathcal{E}_{\text{RSP}}(\rho_{\bold{s}_{0}})=\rho_{\bold{s}}$. Then, $\mathcal{E}_{\text{QN}}$ suffers from uncontrollable noise and becomes classical such that $\mathcal{E}_{c}(\rho_{\bold{s}})=\tilde{\rho}_{{c}|\bold{s}}$ according to Eq.~(\ref{rrc}) where $\rho_{\bold{s}_0}=\rho_{\bold{s}}$.

\subsection*{RSP capability}
Suppose a whole experimental process $\mathcal{E}$ in networks concerning a target process $\mathcal{E}_{\text{RSP}}$ or $\mathcal{E}_{\text{QN}}\circ\mathcal{E}_{\text{RSP}}$ possesses a process matrix beyond the classical networking RSP process matrix's description (\ref{pct}). This description also implies that $\mathcal{E}$ can output states indescribable by the form of Eq.~(\ref{rrc}) or Eq.~(\ref{RSPQNn}) for all input states $\rho_{\bold{s}_{0}}$. In that case, we assert that this process has a type of quantum resource, which we refer to as RSP capability, enabling the implementation of nonclassical state preparation and transmission in a quantum network. Since such a quantum network's state preparation and transmission are nonclassical, we call it a genuine quantum network. We use this term to distinguish the quantum networks derived from other quantum characteristics that cannot guarantee the existence of nonclassical RSP, such as quantum discord and EPR steering.

Notably, the RSP capability describes a quantum network's whole characteristics. It is a resulting quantum property from all the static and dynamic resources and cannot represented by any classical means. Therefore, this quantum resource can validate all the static and dynamic elements, such as the shared pairs, QRs, and quantum channels, utilized by all the networking participants for $\mathcal{E}_{\text{RSP}}$ or $\mathcal{E}_{\text{QN}}\circ\mathcal{E}_{\text{RSP}}$ (Fig.~\ref{RSP}). Moreover, while examination of the pre-shared Bell pair, such as measuring Bell inequality violation, helps confirm the quality of the state resources, it cannot guarantee the quantum characteristics of RSP as a whole because the sender and receiver's operations still need to be examined. Thus, RSP capability provides an active method to examine networks based on RSP. It differs from any statical quantum resources such as superposition or entanglement.

To perform close to the ideal RSP and better than the classical emulation of $\mathcal{E}_c$, satisfying the quantum operations requirements of positive Hermitian process matrix \cite{Nielsen00}, $\tilde{\chi}_{\mathcal{E}}$ (\ref{pt}), becomes necessary for the given experimental process $\mathcal{E}$. Furthermore, to evaluate how $\mathcal{E}$ with a valid process matrix surpasses the classical networking RSP model, $\tilde{\chi}_{\mathcal{E}_c}$ (\ref{pct}), we propose the following criteria for evaluating the static and dynamic combined quantum resources as the RSP capability of $\mathcal{E}$ to achieve genuine quantum networks:

\begin{figure}
\includegraphics[width=7.9cm]{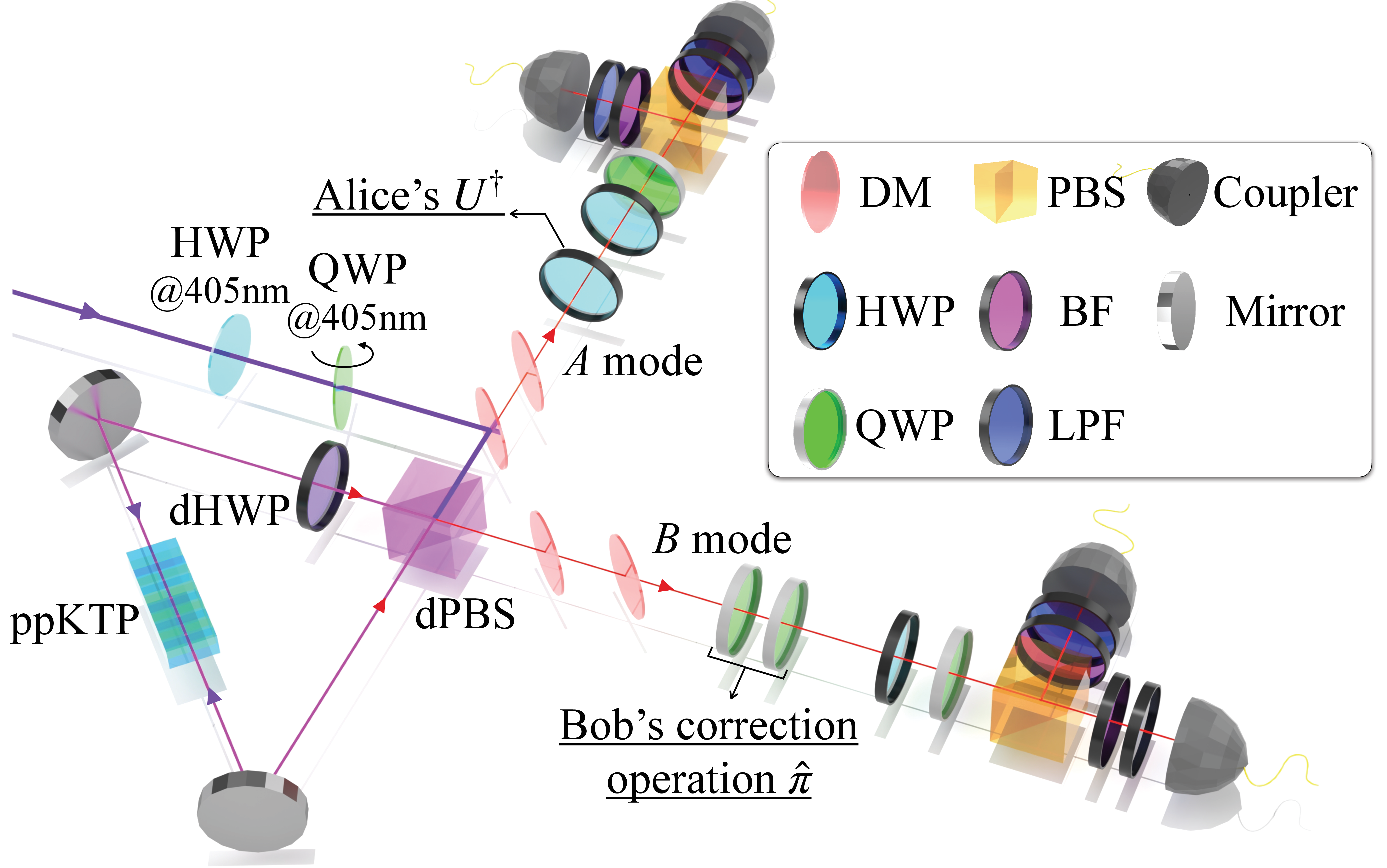}
\caption{Experimental setup for remote state preparation (RSP). A CW diode laser emits a 405 nm beam with a power of 1 mW into a polarization Sagnac interferometer (PSI) consisting of a periodically polled KTP (ppKTP) crystal, a dual-wavelength polarizing beam splitter ($\text {dPBS}$, 405/810 nm), a dual-wavelength half-wave plate ($\text {dHWP}$) set at $\pi/4$, and two mirrors. The horizontal-polarization beam bidirectionally pumps the ppKTP crystal. A type-II down-conversion in each direction generates two orthogonal polarization photons with a wavelength of 810 nm. When the two oppositely propagating photon pairs coherently interfere at $\text {dPBS}$, the joint state of the photon in mode $A$ and photon in mode $B$ becomes a superposition of the states $\ket{H}_{A}\ket{V}_{B}$ and $\ket{V}_{A}\ket{H}_{B}$. By properly setting the angle of the $\text{HWP}$ (405 nm) and rotating the quarter-wave plate ($\text{QWP}$, 405 nm) for the input pump, the photon pairs generated from the PSI are close to the target state $\ket{\Psi^{-}}$, where the logical qubit is encoded as $\ket{H}_k\equiv\ket{0}_3$ and $\ket{V}_k\equiv\ket{1}_3$, for the horizontal ($H$) and vertical ($V$) polarization states of mode $k$, respectively. The photon pairs are analyzed using the HWP, QWP, and PBS and are then collected by couplers with narrowband filters (BFs) and long-pass filters (LPFs) for subsequent detection by single photon counting modules (not shown). Alice’s operation $U^\dag=R^{\dag}(\phi)$ and Bob’s correction operations are implemented by a HWP and two QWPs, respectively (see the Methods and Supplementary Note 2).}
\label{setup}
\end{figure}

\noindent (i) RSP composition $\alpha$: An experimental process $\mathcal{E}$ can quantitatively consist of a linear combination of the classical process $\mathcal{E}_c$ and process that the classical processes cannot represent, $\mathcal{E}_{Q}$. We describe this interpretation in the following form:
\begin{equation}
\chi_{\mathcal{E}}=a\chi_{\mathcal{E}_{Q}}+(1-a)\chi_{\mathcal{E}_c},\label{a}
\end{equation}
where $a\ge 0$ describes the composition ratio of the two processes, and $\chi_{\mathcal{K}}$ for $\mathcal{K}=\mathcal{E},\mathcal{E}_Q,\mathcal{E}_c$ are the normalized process matrices \cite{Nielsen00} of $\mathcal{K}$. We define the RSP composition as the following:
\begin{equation}
\alpha\equiv\min_{\chi_{\mathcal{E}_c}}a,\label{alpha}
\end{equation}
representing the minimum proportion of $\mathcal{E}$ as being the quantum operation $\mathcal{E}_Q$ that $\mathcal{E}_c$ cannot represent,

\noindent (ii) RSP robustness $\beta$: An experimental process $\mathcal{E}$ can become classical by adding noise. We express this interpretation in the following form:
\begin{equation}
\frac{\chi_\mathcal{E}+b\chi_{\mathcal{E}_{\text{noise}}}}{1+b}=\chi_{\mathcal{E}_c},\label{b}
\end{equation}
where  $b \ge 0$ and $\chi_{\mathcal{E}_{\text{noise}}}$ is the porcess matrix of the noise process $\mathcal{E}_{\text{noise}}$. It is worth noting that $\chi_{\mathcal{E}_{\text{noise}}}$ can be arbitrary positive Hermitian for a valid process and is not limited to a specific noise type. We define the RSP robustness as the following: 
\begin{equation}
\beta\equiv\min_{\chi_{\mathcal{E}_c}}b,\label{beta}
\end{equation}
which quantifies the minimum amount of noise $\mathcal{E}_{\text{noise}}$ in the process matrix $\chi_{\mathcal{E}_{\text{noise}}}$ that must be added to make $\mathcal{E}$ classical.

\noindent (iii) Average-state-fidelity criterion for nonclassical networking RSP:
\begin{equation}
\bar{F}_{s}(\mathcal{E})\!\equiv\!\int \!\!d\bold{s}_{0}\!\bra{\bold{s}_{0}}\!U^{\dag}\!\mathcal{E}(\rho_{\bold{s}_{0}})U\!\ket{\bold{s}_{0}}>\bar{F}_{sc},\label{Fcc}
\end{equation}
where 
\begin{equation}
\bar{F}_{sc}\equiv\max_{\mathcal{E}_{c}}\bar{F}_{s}(\mathcal{E}_{c})\sim78.9\%,\label{Fc}
\end{equation}
and the integral is taken over the uniform measure $d\bold{s}_{0}$ and $\int\!d\bold{s}_{0}=1$. See the details of the definition of average state fidelity in Ref.~\cite{Gilchrist05}.

\begin{figure}
\includegraphics[width=8.2cm]{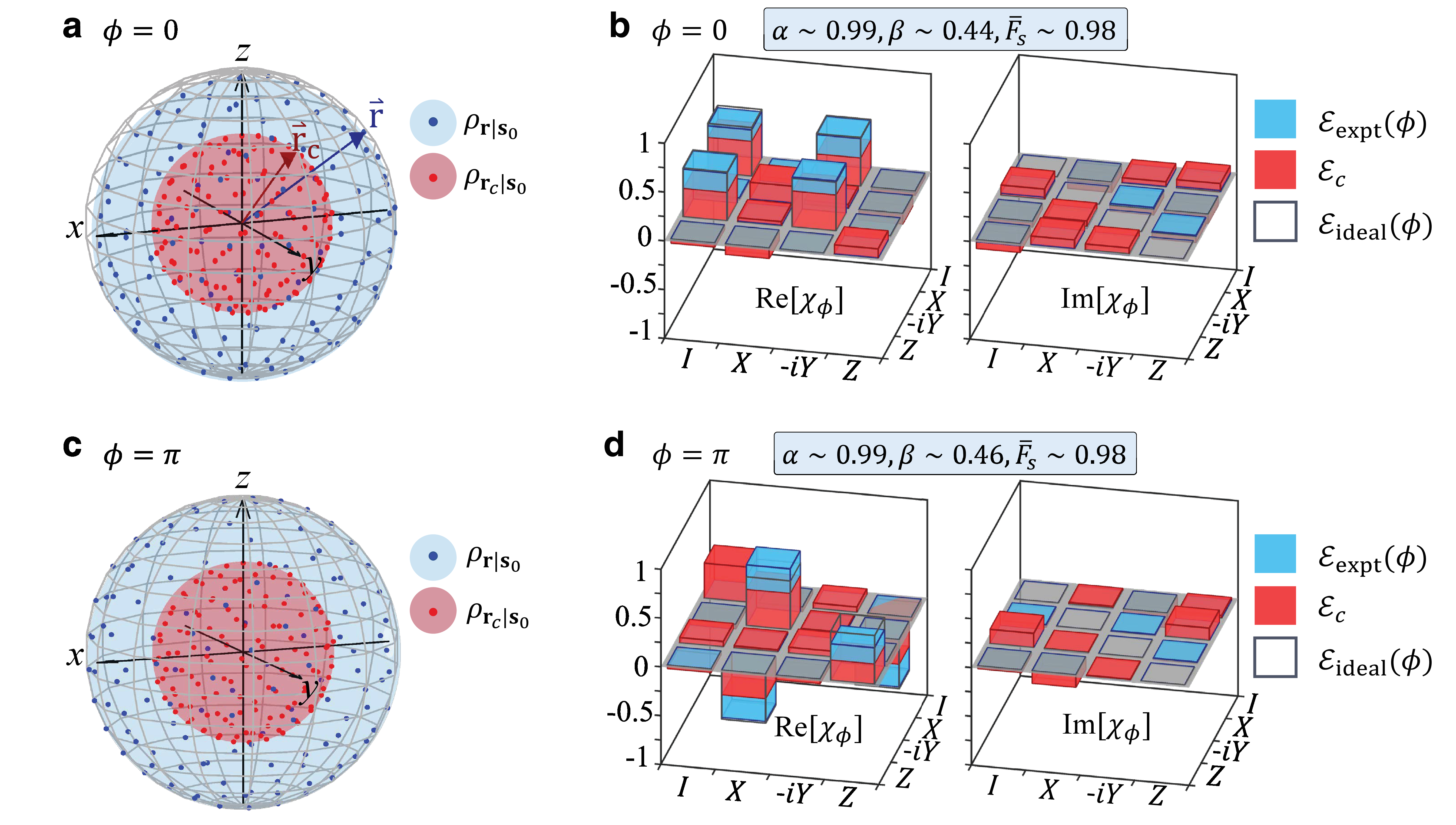}
\caption{Measurements of remote state preparation (RSP) capability. We experimentally quantify the RSP capability by $\alpha$, $\beta$, and $\bar{F}_{s}(\mathcal{E}_\textrm{expt}(\phi))$ to reveal nonclassical RSPs of the implemented $\mathcal{E}_\textrm{expt}(\phi)$ for \textbf{a},\textbf{b} $\phi=0$ and \textbf{c},\textbf{d} $\phi=\pi$. \textbf{a},\textbf{c} The normalized states $\mathcal{E}_\textrm{expt}(\rho_{\bold{s}_{0}})=\rho_{\bold{r}|\bold{s}_{0}}$ shown on the Bloch sphere can visualize the RSP capability. Conditioned on $98$ randomly chosen $\rho_{\bold{s}_{0}}$, we observe all the corresponding output states (blue points) of $\mathcal{E}_\textrm{expt}(\phi)$ having larger Bloch vectors, $|\vec{\bold{r}}|>|\vec{\bold{r}}_{c}|$, compared to the normalized states $\rho_{{c}|\bold{s}_{0}}=\rho_{\bold{r}_{c}|\bold{s}_{0}}$ (red points) derived from the classical RSP $\mathcal{E}_{c}$ with the best process fidelity. Their average vector length better than $0.96$ is superior to the classical limit. The states prepared by the ideal $\mathcal{E}_{\textrm{ideal}}(\phi)$ lie on the surface of the Bloch sphere (gray mesh). \textbf{b},\textbf{d} The normalized process matrices of $\mathcal{E}_\textrm{expt}(\phi)$: $\chi_{\phi}$, consisting of the real (Re) and imaginary (Im) parts, are measured (blue bars) and used to derive $\alpha$, $\beta$, and $\bar{F}_{s}$; from which the output states $\rho_{\bold{r}|\bold{s}_{0}}$ can then be determined as well for the input states $\rho_{\bold{s}_{0}}$ (see Supplementary Note 1). They are compared to the process matrices for $\mathcal{E}_c$ (red bars) and those for $\mathcal{E}_{\textrm{ideal}}(\phi)$ (gray frames).}
\label{UT}
\end{figure}

In measuring the composition $\alpha$ (\ref{alpha}) and robustness $\beta$ (\ref{beta}) and determining the threshold $\bar{F}_{sc}$~(\ref{Fc}), we can express the optimization problem concerning the classical process $\mathcal{E}_{c}$ in the standard semidefinite programming (SDP) problem \cite{Cavalcanti17} and solve the problem with a numerical SDP solver \cite{Lofberg,sdpsolver} (see the Methods section). For example, $\mathcal{E}_{\text{RSP}}$ and $\mathcal{E}_{\text{QN}}\circ\mathcal{E}_{\text{RSP}}$ are nonclassical in the ideal networking RSP  for arbitrary $U$ and have $\alpha=1$, $\beta=0.464$, and $\bar{F}_{s}=1$. For comparison, the first criterion $\alpha$ (\ref{alpha}) regarding the composition concretely determines the maximum proportion of the classical process $\mathcal{E}_c$ in an experimental process $\mathcal{E}$. The second criterion of robustness $\beta$ (\ref{beta}) describes how close the experimental process $\mathcal{E}$ is to the classical process $\mathcal{E}_c$. Such differences between their definitions provide different insights into quantifying the RSP capability. Therefore, they represent two methods to quantify the RSP capability in experiments. As shown in the experimental demonstration below, these values decrease due to noise contamination.

The RSP capability enables a payoff for nonclassical state preparation and transmission. Focusing on the utility of state preparation, examining the RSP capability in a practical RSP now allows uncharacterized (or uncalibrated) shared pairs, Alice’s qubit manipulations, and Bob’s corrections for experimental RSPs. Thus, examining the RSP capability helps confirm the resulting genuine quantum networks, which cannot be achieved using the existing verification methods \cite{Dakic10,Luo10,Killoran10}. For state transmission, the RSP capability helps validate the ability of $\mathcal{E}_{\text{QN}}$ to preserve prepared states in the network and rules out the cases in Eq.~(\ref{RSPQNn}). Furthermore, the classical model [Eqs.~(\ref{pct},\ref{rrc})] emulates all the hardware and resource necessities required to realize networking RSP. Therefore, the RSP capability provides a dynamic resource compared with static state characteristics such as EPR steering \cite{Wiseman07} and its temporal analogs \cite{Chen14,Chen16}. Moreover, Eq.~(\ref{pct}) uniquely considers concrete emulations of the measurement-based RSP protocol and hence differs from the existing classical input-output model used for identifying teleportation \cite{Chen2021}. This feature also makes the criteria (\ref{alpha},\ref{beta},\ref{Fc}) distinct from those for teleportation~\cite{Chen2021}.

\subsection*{Experimental demonstration of the networking RSP process}
In our experimental demonstration of the networking RSP process, we created a bidirectionally pumped down-conversion source of polarization Sagnac interferometer (PSI) using a periodically polled KTP (ppKTP) crystal to generate polarization-entangled photon pairs \cite{Kim06,Chaisson22} (see Fig.~\ref{setup}). We observed $\sim\!\!1000$ photon pairs per second and measured the state fidelity of $F\equiv\text{tr}(\rho_{\text{expt}}\!\ket{\Psi^{-}}\!\!\bra{\Psi^{-}})=(98.62 \pm 0.26)\%$ between the created state $\rho_{\text{expt}}$ and the target state $\ket{\Psi^{-}}$ for RSP. See Supplementary Note 2 for the measured density matrix $\rho_{\text{expt}}$. We then implemented the RSP protocol with a high-quality EPR-pair source and precise realization of the operation:
\begin{equation}
U^\dag=R^{\dag}(\phi)=\frac{1}{\sqrt{2}}(\ket{0}_{\!33}\!\!\bra{0}\!+\!e^{-i\phi}\ket{0}_{\!33}\!\!\bra{1}\!+\!\ket{1}_{\!33}\!\!\bra{0}\!-\!e^{-i\phi}\ket{1}_{\!33}\!\!\bra{1}).\nonumber
\end{equation}
The Methods section details the main steps for implementing the RSP protocol and how to get the experimental process matrix (\ref{pt}).

We first realized two different kinds of RSP, $\mathcal{E}_\textrm{expt}(\phi)$, for $\phi=0,\pi$. For convenience, we denote the ideal RSP process as $\mathcal{E}_{\textrm{ideal}}(\phi)$. Both RSPs are nonclassical and possess RSP capability, as evaluated using the above criteria in (i)-(iii). Figure~\ref{UT} shows the concrete comparisons between $\mathcal{E}_\textrm{expt}(\phi)$, $\mathcal{E}_{\textrm{ideal}}(\phi)$ and $\mathcal{E}_{c}$.

\begin{figure}
\includegraphics[width=7.9cm]{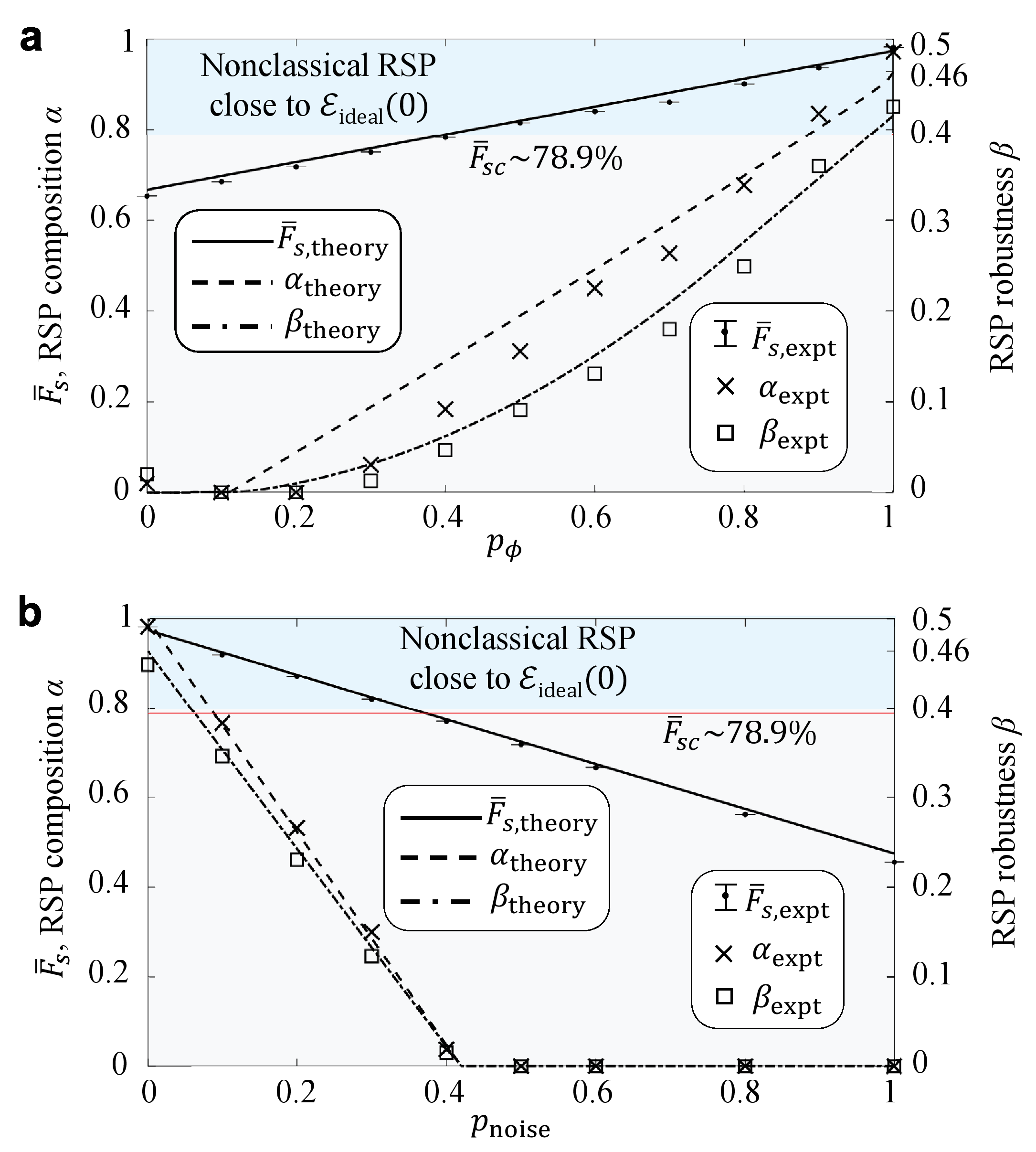}
\caption{Experimental transition between classical and nonclassical remote state preparations (RSPs). We implemented experimental RSPs of $\mathcal{E}_\textrm{expt}(\phi)$ for $\phi=0$ using \textbf{a} the noisy state $\rho(p_{\phi})$ [Eq.~(\ref{noise})] and \textbf{b} the Werner state $\rho_{W}(p_{\text{noise}})$ [Eq.~(\ref{wstate})]. The experimental RSP capabilities, $\alpha_{\text{expt}}$ and $\beta_{\text{expt}}$, and the average state fidelities, $\bar{F}_{s,\text{expt}}$, are highly consistent with the theoretical predictions of $\alpha_{\text{theory}}$, $\beta_{\text{theory}}$, and $\bar{F}_{s,\text{theory}}$, respectively, derived from $\rho(p_{\phi})$ for \textbf{a} and $\rho_{W}(p_{\text{noise}})$ for \textbf{b} (see Supplementary Note 3). An RSP process is nonclassical and close to $\mathcal{E}_\textrm{ideal}(0)$ only when satisfying the criterion~(\ref{Fc}). For example, while the created photon pairs at $p_{\phi}=0.4$ possess quantum discord \cite{Dakic10} ($\mathcal{D}_{\text{theory}}\sim0.075$, $\mathcal{D}_{\text{expt}}\sim0.054$) and steerability regarding the steerable weight \cite{Skrzypczyk2014} ($SW_{\text{theory}}\sim0.183$, $SW_{\text{expt}}\sim0.095$) (see Supplementary Note 4), the resulting RSP is not superior to $\mathcal{E}_{c}$ for $\bar{F}_{s,\text{theory}}\sim 78.62\%$ and $\bar{F}_{s,\text{expt}}\!=\!\!(78.27\pm0.03)\%$. Moreover, theoretically and experimentally, an RSP process is nonclassical only when the noise is lower than a certain threshold. An experimental RSP can be a classical process with $\alpha=0$ and $\beta=0$ even though the state of the underlying photon pairs $\rho_{W}(p_{\text{noise}})$ possesses quantum discord, e.g., ($\mathcal{D}_{\text{theory}}\!\sim\!0.118$, $\mathcal{D}_{\text{expt}}\!\sim\!0.156$) for $p_{\text{noise}}=0.5$.}
\label{IM}
\end{figure}

In the following, we demonstrate how, under noise contamination, the created high-fidelity EPR pairs of the state $\rho_{\text{expt}}$ become unhelpful in supporting nonclassical RSPs with $\alpha,\beta>0$ and (or) $\bar{F}_{s}>\!\bar{F}_{sc}$. Without loss of generality, we focus on the experimental RSP $\mathcal{E}_\textrm{expt}(\phi)$ for $\phi=0$. We first consider the case where $\rho_{\text{expt}}$ mixes with the noise of an incoherent mixture $\rho_\text{sep}\!=\!(\ket{0}_{\!33}\!\!\bra{0}\!\otimes\!\ket{1}_{\!33}\!\!\bra{1}\!+\!\ket{1}_{\!33}\!\!\bra{1}\!\otimes\!\ket{0}_{\!33}\!\!\bra{0})/2$ as
\begin{equation}
\rho(p_{\phi})=p_{\phi}\rho_{{\rm expt}}+(1-p_{\phi})\rho_\text{sep},\label{noise}
\end{equation}
for $0\leq p_{\phi}\leq 1$. We note that this demonstration equivalently reflects how the interference imperfection at dPBS in PSI (Fig.~\ref{setup}), as quantified by $1-p_{\phi}$, affects the RSP capability of the whole photonic system (see Fig.~\ref{IM}\textbf{a}). It is worth noting that an RSP that uses photon pairs with non-zero quantum discord and EPR steerability may not satisfy the fidelity criterion (\ref{Fc}). Figure~\ref{IM}\textbf{b} describes the experimental RSPs using the Werner state of the following form
\begin{equation}
\rho_{W}(p_{\text{noise}})=(1-p_{\text{noise}})\rho_{{\rm expt}}+\frac{p_{\text{noise}}}{4}I\otimes I,\label{wstate}
\end{equation}
for $0\leq p_{\text{noise}}\leq 1$. The classical process $\mathcal{E}_{c}$ can even describe specific RSP processes that rely on quantum discord. The above experimental demonstration clearly shows that quantum discord and EPR steering cannot guarantee the existence of nonclassical RSP. This also implies that the RSP-based genuine quantum networks exist only when an experimental process has the RSP capability.

\section*{conclusions}
We have introduced a quantum resource that we call the RSP capability, and have shown that it is the necessary resource that enables a networking RSP to outperform classical processes of preparing and transmitting states in networks without the need for entanglement and qubit unitaries. This new resource specifies all the static and dynamic elements required in nonclassical networking RSP experiments for genuine quantum networks. This cannot be achieved using the existing RSP verification methods \cite{Dakic10,Luo10,Killoran10}, which do not consider classical physics into account. We have demonstrated experimentally how the RSP capability of a photonic system can be measured using three criteria (i)-(iii) for different types of polarization-correlated photon states created from a Sagnac interferometer. We reveal that while polarization-correlated photon pairs possess quantum discord, the classical model still can describe the corresponding resulting state preparations. Therefore, our results show that nonclassically preparing remote states is significant for genuine quantum networks and essential for applications where ruling out any classical simulations of networking RSP is necessary \cite{Jeffrey04,Barreiro10,Bao12,Gustiani2021}. They may provide insight into identifying nonclassical processes for general measurement-based quantum-information processing \cite{raussendorf2001,briegel2009,you2007,tanamoto2009,wang2010,Chou18,Pirker18}.

As summarized in the introduction, RSP is necessary to organize a quantum memory state for quantum-memory-related applications. Therefore, it may be interesting to investigate further the utility of the RSP capability as a resource for memory-assisted measurement-device-independent quantum key distribution \cite{Panayi14}, global quantum networking \cite{Gundo21,Walln22,Gundo23}, and networked quantum technologies~\cite{Belenchia22}. It may help deliver insights into various tradeoffs in performance, such as link availability, and range vs loss. According to the basic concept and method of the RSP protocol, it is easy to find a similarity between the RSP and the entanglement-based QKD, where the sender helps establish useful information for the receiver in both cases and creates a connection between them. Therefore, this preliminary finding makes it attractive to investigate whether our formalism of identifying genuine quantum networks has a deeper relationship with the security analysis utilized in the entanglement-based QKD.

\section*{Methods}

\subsection*{Implementation of the RSP protocol}

We implemented the RSP protocol according to the descriptions in the Results section on the RSP's quantum operation. We first realized the required operation $U^\dag=R^\dag\!(\phi)$ in the first step of the RSP protocol with the optical setup shown in Fig.~\ref{setup}, where the rotation operator is $R(\phi)=(\ket{0}_{\!33}\!\!\bra{0}\!+\!e^{i\phi}\ket{1}_{\!33}\!\!\bra{0}\!+\!\ket{0}_{\!33}\!\!\bra{1}\!-\!e^{i\phi}\ket{1}_{\!33}\!\!\bra{1})\!/\!\sqrt{2}$. First, for $R(0)$, we set the HWP, which is placed before the polarization analyzer in mode $A$, at $22.5^{\circ}$ to realize the operation $U^{\dag}$ required on Alice's side. In the ideal case, this makes the state $\ket{\Psi^{-}}$ become
\begin{equation}
U^\dag\ket{\Psi^{-}}=\frac{1}{\sqrt{2}}(\ket{\bold{s}_{0}}\otimes R(0)\ket{\bold{s}_{0}^{\bot}}-\ket{\bold{s}_{0}^{\bot}}\otimes R(0)\ket{\bold{s}_{0}}).\label{explain1}
\end{equation}
Then, we measure Alice's qubit in the basis $\{\ket{\bold{s}_{0}},\ket{\bold{s}_{0}^{\bot}}\}=\{\ket{0}_m,\ket{1}_m\}$ by the polarization analyzer in mode $A$, where $\ket{n}_{m}$ for $n=0,1$ are the eigenvectors of the Pauli matrices $\sigma_{m}$ for $m=1,2,3$. Experimentally, the observables $\sigma_{m}$ are measured by setting the HWP and the QWP of the polarization analyzer at suitable angles. For example, when we set the optical axes of the HWP and the QWP as $22.5^{\circ}$ and $0^{\circ}$, respectively, with respect to the vertical axis, denoted as $\left(22.5^{\circ},0^{\circ} \right)$, the polarization analyzer can measure $\sigma_{1}$, i.e., measuring Alice's qubit in the basis of $\{\ket{0}_1,\ket{1}_1\}$. Similarly, $\sigma_{2}$ and $\sigma_{3}$ are measured under the angle settings: $\left( 0^{\circ}, 45^{\circ} \right)$ and $\left( 0^{\circ}, 0^{\circ} \right)$, respectively. Then, a subsequent operation of the PBS in the polarization analyzer helps distinguish the states $\ket{0}_{3}$ and $\ket{1}_{3}$, which completes Alice's measurement.

Next, we realized Bob's correction operations on his qubit according to Alice's measurement results. When Alice measures her qubit in the basis $\{\ket{\bold{s}_{0}}=\ket{0}_3,\ket{\bold{s}_{0}^{\bot}}=\ket{1}_3\}$ according to the experimental settings described in the last paragraph, her measurement results imply the following Bob's qubit states in the ideal case of Eq.~(\ref{explain1}): $\ket{0}_3\rightarrow R(0)\ket{1}_3$ and $\ket{1}_3\rightarrow R(0)\ket{0}_3$. Therefore, according to the $\ket{0}_3$ measurement result of Alice's qubit, we set the two QWPs as $0^{\circ}$ on Bob's side to serve as a $\hat{\pi}$ operation on Bob's state $R(0)\ket{1}_3$. Idealy, after this correction, his state becomes $\hat{\pi}R(0)\ket{1}_3=R(0)\ket{0}_3$. Whereas, when Alice's result is $\ket{1}_3$, Bob's state is already in a correct state $R(0)\ket{0}_3$; we did not apply the correction operation to the qubit.

Moreover, our RSP experiments are passive. Alice's measurement outcomes are not communicated in real-time to Bob, and the conditional unitaries for qubit corrections are passively performed on the output state. Our passive RSP experiments consist of events with and without corrections. For the events without corrections, like the heralded RSP discussed in the Results section, Alice's result $\ket{1}_3$ as a heralded signal shows that we have prepared Bob's qubit as $R(0)\ket{0}_3$. For the events with corrections, we set the two QWPs on Bob's side as the correction mode before Alice's measurements. We then get a prepared state $R(0)\ket{0}_3$ when the state $\ket{0}_3$ on Alice's side is detected, which acts as a heralded signal.

In analyzing our experimental implementations, the realized rotation and correction operators are described by quantum operations $\mathcal{E}_{R(0)}$ and $\mathcal{E}_{\hat{\pi}}$, respectively. As a result, compared to the ideal RSP (\ref{ideal}), the resulting prepared state is
\begin{equation}
\mathcal{E}(\ket{0}_{\!33}\!\bra{0})=p_A(1)\mathcal{E}_{R(0)} \left(\ket{0}_{\!33}\!\bra{0}\right)+p_A(0)\mathcal{E}_{\hat{\pi}}\circ\mathcal{E}_{R(0)} \left(\ket{1}_{\!33}\!\bra{1}\right),\label{exp}
\end{equation}
where $p_A(n)$ for $n=0,1$ denotes the probabilities that Alice obtains the results $n$. The probabilities $p_A(n)$ can also be considered the probabilities of the events with and without corrections appearing in our passive RSP experiments described above. The output state of the experimental RSP, $\mathcal{E}(\ket{0}_{\!33}\!\bra{0})$, is an average of these two events' resulting states, $\mathcal{E}_{R(0)} \left(\ket{0}_{\!33}\!\bra{0}\right)$ and $\mathcal{E}_{\hat{\pi}}\circ\mathcal{E}_{R(0)} \left(\ket{1}_{\!33}\!\bra{1}\right)$. For the ideal case, Eq.~(\ref{exp}) becomes $\mathcal{E}(\ket{0}_{\!33}\!\bra{0})=U\ket{0}_{\!33}\!\bra{0}U^\dag$, as described by Eqs.~(\ref{ideal}) and (\ref{RSPQO}) where $p_A(n)=\tr(\ketbra{\Psi^{-}}\hat{M}_A)=1/2$. Similarly, the above experimental two-step RSP applies to $U=R(\pi)$ and other input states, such as the other eigenstates of the Pauli matrices, $\ket{n}_{m}$.

\subsection*{Experimental process matrix $\tilde{\chi}_{\mathcal{E}}$}

To get the process matrix $\tilde{\chi}_{\mathcal{E}}$ (\ref{pt}) of an experimental RSP process $\mathcal{E}$ according to the process tomography of quantum operations formalism \cite{Nielsen00}, we first perform state tomography \cite{Nielsen00} to obtain the density matrix of the output state $\mathcal{E}(\ket{n}_{mm}\!\!\bra{n})$ by the polarization analyzer in mode $B$. The required measurement settings for the observables $\sigma_{m}$ are the same as the settings of the polarization analyzer in mode $A$, as described in the first step of the RSP implementation in the above section. Let us take $\ket{\bold{s}_{0}}=\ket{0}_3$ and $\ket{\bold{s}_{0}^{\bot}}=\ket{1}_3$ for example; our goal is to obtain the density matrix of $\mathcal{E}(\ket{0}_{\!33}\!\bra{0})$ (\ref{exp}). We perform the measurement of $\sigma_{1}$, $\sigma_{2}$, and $\sigma_{3}$ on $\mathcal{E}(\ket{0}_{\!33}\!\bra{0})$ in mode $B$ conditioned on Alice's results of measurements in mode $A$. Therefore, to determine the density matrix of $\mathcal{E}(\ket{0}_{33}\!\bra{0})$, we measure $p_A(n)$ and tomographically get the states $\mathcal{E}_{R(0)} \left(\ket{0}_{\!33}\!\bra{0}\right)$ and $\mathcal{E}_{\hat{\pi}}\circ\mathcal{E}_{R(0)} \left(\ket{1}_{\!33}\!\bra{1}\right)$. With the experimentally determined density operators $\mathcal{E}(\ket{n}_{mm}\!\!\bra{n})$, we get the process matrix $\tilde{\chi}_{\mathcal{E}}$ (\ref{pt}) of $\mathcal{E}$ according to the process tomography protocol \cite{Nielsen00} and from which we obtain $\alpha$, $\beta$, and $\bar{F}_{s}$.

\subsection*{Semidefinite programming}
We first describe the methods we used to cast the optimization problems of $\alpha$ (\ref{alpha}), $\beta$ (\ref{beta}), and $\bar{F}_{sc}$ (\ref{Fc}) as SDP problems by formulating the introduced equations to the standard SDP optimization problem. A standard SDP problem aims to minimize a real linear objective function with a set of optimization parameters, and the constraints it satisfies can be represented as a set of linear matrix inequality constraints. Calculating the RSP capability of a given RSP process $\mathcal{E}$ with a process matrix, $\alpha$ (\ref{alpha}) or $\beta$ (\ref{beta}), can be viewed as the objective function we have to minimize. The choice of the parameters $\chi_{\mathcal{E}_c}$, $\rho_{\lambda}$, and $p(\lambda)$ in the classical networking RSP model from  Eq.~(\ref{ec}) are the optimization parameters. Moreover, all the process and density matrices in Eqs.~(\ref{alpha},\ref{beta}) should be positive semidefinite, i.e., satisfy the following linear matrix inequality constraints: ${\chi}_{\mathcal{E}_c}\geq0$ and ${\rho}_{\lambda}\geq 0, \forall \lambda$. Thus, we cast the optimization of $\alpha$ (\ref{alpha}) and $\beta$ (\ref{beta}) as SDP problems. Similarly, for a given target RSP $\mathcal{E}_{\text{RSP}}$ with the process matrix $\chi_{\text{RSP}}$, the fidelity bound (\ref{Fc}) can be solved as an SDP problem since process fidelity is a real linear objective function. Our methods outlined above are similar to the existing work in Ref.~\cite{Cavalcanti17}, where they use density matrices as optimization parameters in SDP in the steering quantifiers.

To detail the optimization of the objective functions $\alpha$, $\beta$, and $\bar{F}_{sc}$ via SDP \cite{Lofberg,sdpsolver,Cavalcanti17}, we first expressed the optimization parameters in an unnormalized process matrix $\tilde{\chi}_{\mathcal{E}_c}$ consisting of eight unnormalized matrices $\tilde{\rho}_{\lambda}$, where $\tilde{\rho}_{\lambda}=2p(\lambda)\rho_{\lambda}$ [see Eqs.~(\ref{ecall},\ref{pct})], $\tilde{\chi}_{\mathcal{E}_c}=(1-a)\chi_{\mathcal{E}_c}$ for obtaining $\alpha$ in Eq.~(\ref{a}), and $\tilde{\chi}_{\mathcal{E}_c}=(1+b)\chi_{\mathcal{E}_c}$ for obtaining $\beta$ in Eq.~(\ref{b}). It is worth noting that the number of optimization parameters depends on the computation difficulty of an SDP. The optimization parameters in our SDP optimization problem are the eight parameters, $\tilde{\rho}_{\lambda}$. The six output states in Eq.~(\ref{ecall}) for constructing classical networking RSP process matrix can be expressed by the optimization parameters $\tilde{\rho}_{\lambda}$, i.e.,
\begin{eqnarray}
&&\tilde{\rho}_{c|v_{01}}=\sum_{\lambda=1,2,3,4}\tilde{\rho}_{\lambda},\ \tilde{\rho}_{c|v_{11}}=\sum_{\lambda=5,6,7,8}\tilde{\rho}_{\lambda},\ \nonumber\\
&&\tilde{\rho}_{c|v_{02}}=\sum_{\lambda=1,2,5,6}\tilde{\rho}_{\lambda},\ \tilde{\rho}_{c|v_{12}}=\sum_{\lambda=3,4,7,8}\tilde{\rho}_{\lambda},\ \nonumber\\
&&\tilde{\rho}_{c|v_{03}}=\sum_{\lambda=1,3,5,7}\tilde{\rho}_{\lambda},\ \tilde{\rho}_{c|v_{13}}=\sum_{\lambda=2,4,6,8}\tilde{\rho}_{\lambda},\nonumber
\end{eqnarray}
where $\text{tr}(\tilde{\chi}_{\mathcal{E}_c})=\sum_{\lambda}\text{tr}(\tilde{\rho}_{\lambda})$.
The process matrix of the classical process $\tilde{\chi}_{\mathcal{E}_c}$ must satisfy the definition of process matrices of being positive semidefinite with the following constraints: $\tilde{\chi}_{\mathcal{E}_c}\geq0$, and the density matrices should satisfy $\tilde{\rho}_{\lambda}\geq 0, \forall \lambda$. According to the mathematical structure of the objective functions $\alpha$ and $\beta$, the optimization for the RSP capabilities is exact for all valid processes $\mathcal{E}$. There is no duality gap since, for cases where the classical RSP model can describe a process, the primal optimal values of $\alpha$ and $\beta$ are zero. If, on the other hand, the classical RSP model cannot describe the process, the primal optimal value will be a finite value bounded by the SDP constraints for obtaining the RSP capabilities $\alpha$ and $\beta$. Similarly, the optimization for the fidelity bound $\bar{F}_{sc}$ is exact for ideal networking RSP, since, according to the definition of fidelity, the primal optimal value is bounded by the SDP constraints for $\bar{F}_{sc}$ and is a finite value from zero to one.

With the above descriptions of the optimization parameters and constraints for SDP, we respectively re-express Eqs.~(\ref{alpha}), (\ref{beta}), and (\ref{Fc}) in SDP formulations as follows.\\
\noindent (i) Determination of $\alpha$: The RSP composition $\alpha$ can be obtained by minimizing the following objective function via SDP with MATLAB \cite{code}:
\begin{subequations}
	\begin{align}
       \text{given} &\quad\chi_{\mathcal{E}} \\
       \alpha = \min_{\tilde{\chi}_{\mathcal{E}_c}} &\quad 1-\text{tr}(\tilde{\chi}_{\mathcal{E}_c})\\
		\text{subject to}&\quad \tilde{\rho}_{\lambda}\geq 0,& &\forall \lambda, \\
		&\quad \tilde{\chi}_{\mathcal{E}_c}\geq 0, \\
        &\quad \text{tr}(\tilde{\rho}_{c|v_{0m}})=\text{tr}(\tilde{\rho}_{c|v_{1m}}),& &\forall m, \\
		&\quad \chi_{\mathcal{E}}-\tilde{\chi}_{\mathcal{E}_c}\geq 0,
	\end{align}
\end{subequations}
where $\tilde{\chi}_{\mathcal{E}_c}=(1-a)\chi_{\mathcal{E}_c}$. The first constraint ensures that all possible remote states $\tilde{\rho}_{\lambda}$ in the classical RSP model are positive semidefinite. The second constraint ensures that the $\tilde{\chi}_{\mathcal{E}_c}$ must be positive semidefinite since it is a valid process matrix, and this constraint makes $\alpha\leq 1$. The third constraint ensures that, from the $\tilde{\chi}_{\mathcal{E}_c}$, the observed proportion of each input state is the same in the three bases. The fourth constraint makes the process matrix of $\mathcal{E}_{Q}$ satisfy the definition of process matrices which need to be positive semidefinite and thus leads to $\alpha\geq 0$.\\

\noindent (ii) Determination of $\beta$: The RSP robustness $\beta$ of $\chi_{\mathcal{E}}$ can be obtained by using SDP to solve the objective function with MATLAB \cite{code} by:
\begin{subequations}
	\begin{align}
       \text{given} &\quad\chi_{\mathcal{E}} \\
		\beta = \min_{\tilde{\chi}_{\mathcal{E}_c}} &\quad \text{tr}(\tilde{\chi}_{\mathcal{E}_c})-1\\
		\text{subject to}&\quad \tilde{\rho}_{\lambda}\geq 0,& &\forall \lambda, \\
		&\quad \tilde{\chi}_{\mathcal{E}_c}\geq 0, \\
        &\quad \text{tr}(\tilde{\rho}_{c|v_{0m}})=\text{tr}(\tilde{\rho}_{c|v_{1m}}),& &\forall m, \\
        &\quad \tilde{\chi}_{\mathcal{E}_c}-\chi_{\mathcal{E}}\geq 0, \\
		&\quad \text{tr}(\tilde{\chi}_{\mathcal{E}_c})\geq 1,
	\end{align}
\end{subequations}
where $\tilde{\chi}_{\mathcal{E}_c}=(1+b)\chi_{\mathcal{E}_c}$. The first, second, and third constraints are the same as those used to calculate $\alpha$. The fourth constraint is the condition of positive semidefiniteness for noise process $b\chi_{\mathcal{E}_{\text{noise}}}=(1+b)\chi_{\mathcal{E}_c}-\chi_{\mathcal{E}}$ in Eq.~(\ref{b}). The last constraint ensures that $\beta\geq0$.\\

\noindent (iii) Determination of $\bar{F}_{sc}$: With the process matrix $\chi_{\text{RSP}}$ of $\mathcal{E}_{\text{RSP}}$, the process fidelity upper bound of $\tilde{\chi}_{\mathcal{E}_c}$ and $\chi_{\text{RSP}}$ can be solved through maximizing the following objective function with MATLAB \cite{code}, 
\begin{subequations}
	\begin{align}
       \text{given} &\quad\chi_{\text{RSP}} \\
		F_{\mathcal{E}_{c}} = \max_{\tilde{\chi}_{\mathcal{E}_c}} &\quad \text{tr}(\tilde{\chi}_{\mathcal{E}_c}\chi_{\text{RSP}})\\
		\text{subject to}&\quad \tilde{\rho}_{\lambda}\geq 0,& &\forall \lambda, \\
        &\quad \tilde{\chi}_{\mathcal{E}_c}\geq 0, \\
        &\quad \text{tr}(\tilde{\rho}_{c|v_{0m}})=\text{tr}(\tilde{\rho}_{c|v_{1m}}),& &\forall m, \\
		&\quad \text{tr}(\tilde{\chi}_{\mathcal{E}_c})= 1,
	\end{align}
\end{subequations}
where the first, second, and third constraints are the same as the constraints used in calculating $\alpha$ and $\beta$, and the third constraint makes the $\tilde{\chi}_{\mathcal{E}_c}$ be a normalized matrix to calculate fidelity. While we get the process fidelity bound $F_{\mathcal{E}_{c}}=0.683$, the average-state-fidelity bound $\bar{F}_{sc}=78.9\%$ can be calculated through $\bar{F}_{sc}=(2F_{\mathcal{E}_{c}}+1)/3$ \cite{Hofmann05,Gilchrist05}.\\

\section*{Acknowledgements}
This work was partially supported by the National Science and Technology Council, Taiwan, under grant nos. MOST 107-2628-M-006-001-MY4, MOST 111-2119-M-007-007, MOST 111-2123-M-006-001, MOST 111-2112-M-006-033, NSTC 112-2112-M-006-029, and NSTC 113-2811-M-006-019.
%



\end{document}


\title{Supplementary Information:\\ Preparing Remote States for Genuine Quantum Networks}

\author{Shih-Hsuan Chen$^{1,2}$}
\author{Chan Hsu$^{1,2}$}
\author{Yu-Chien Kao$^{1,2}$}
\author{Bing-Yuan Lee$^{1,2}$}
\author{Yuan-Sung Liu$^{1,2}$}
\author{Yueh-Nan Chen$^{2,3}$}
\author{Che-Ming Li$^{1,2,4,}$}
\email{cmli@mail.ncku.edu.tw}

\affiliation{$^{1}$Department of Engineering Science, National Cheng Kung University, Tainan 70101, Taiwan}
\affiliation{$^{2}$Center for Quantum Frontiers of Research \& Technology, National Cheng Kung University, Tainan 70101, Taiwan}
\affiliation{$^{3}$Department of Physics, National Cheng Kung University, Tainan 70101, Taiwan}
\affiliation{$^{4}$Center for Quantum Science and Technology, Hsinchu 30013, Taiwan}
\begin{abstract}
This supplementary information provides the required material to support our theoretical and experimental results presented in the main text: We meticulously supplement the quantum process tomography and process matrix for quantum operations requirements to precisely detail how the RSP capability of an experimental process of preparing remote states can be quantified and identified. We describe the state of the experimentally created photon pairs from a polarization Sagnac interferometer in terms of density operator. We then introduce theoretical models to explain how RSP capability varies with the noise of an incoherent mixture and with the noise weight in the Werner state. To compare RSP capability with the discord and EPR steerability, we utilize the geometric discord and steerable weight to measure the quantum discord and EPR steerability of the created photon pairs in our experiment.
\end{abstract}

\maketitle

\section*{Supplementary Note 1: Process tomography and process matrix}

The experimental process $\mathcal{E}$ can be fully described in practical experiments using the process tomography algorithm \cite{Nielsen00,Chuang97} and then represented by a positive Hermitian matrix process matrix $\tilde{\chi}_{\mathcal{E}}$, as described in the main text. Explicitly, the quantum operation $\mathcal{E}$ is related to its process matrix $\tilde{\chi}_{\mathcal{E}}$ in the following relation:
\begin{equation}
\mathcal{E}(\rho_{\bold{s}_{0}})=\sum_{m,m'=0}^3(\tilde{\chi}_{\mathcal{E}})_{mm'}E_m\rho_{\bold{s}_{0}}E^{\dag}_{m'},\label{chi}
\end{equation}
where the matrix elements $(\tilde{\chi}_{\mathcal{E}})_{mm'}$ constitute the $4\times4$ process matrix $\tilde{\chi}_{\mathcal{E}}$, and $E_0=I$, $E_1=X$, $E_1=-iY$, and $E_3=Z$ form a basis for the set of operators on the state space. As the process matrix $\tilde{\chi}_{\mathcal{E}}$ is known, Eq.~(\ref{chi}) can fully describe how an input state evolves. Equation (5) in the main text represents the complete form of the process matrix $\tilde{\chi}_{\mathcal{E}}$. See Refs.~\cite{Nielsen00,Chuang97} for a detailed discussion.

\subsubsection*{Figure 3 in the main text}

In Fig.~3\textbf{a} and 3\textbf{c} in the main text, the 98 output states from $\mathcal{E}=\mathcal{E}_\textrm{expt}(\phi)$ for $\phi=0,\pi$ and those from the classical RSP $\mathcal{E}_{c}$ with the best process fidelity shown on the Bloch sphere are determined through the respective normalized process matrices $\chi_\phi$ and $\chi_{\mathcal{E}_c}$. We measure the two process matrices of $\mathcal{E}_\textrm{expt}(\phi)$ and determine the corresponding $\chi_{\mathcal{E}_c}$ according to the process tomography described above. See Fig.~3\textbf{b} and 3\textbf{d}. As the process matrices of $\mathcal{E}_\textrm{expt}(\phi)$ and $\mathcal{E}_{c}$ are obtained, one can get the output states $\mathcal{E}_\textrm{expt}(\phi)(\rho_{\bold{s}_{0}})= \rho_{\bold{r}|\bold{s}_{0}}$ and $\mathcal{E}_{c}(\rho_{\bold{s}_{0}})= \rho_{\bold{r}_c|\bold{s}_{0}}$ for arbitrary input states $\rho_{\bold{s}_{0}}$. We randomly choose 98 of $\rho_{\bold{s}_{0}}$ and determine the corresponding output states by Eq.~(\ref{chi}). The states $ \rho_{\bold{r}|\bold{s}_{0}}$ and $ \rho_{\bold{r}_c|\bold{s}_{0}}$ are therefore represented by red and blue points on the Bloch sphere, respectively.

\section*{Supplementary Note 2: The Sagnac-based polarization-entangled photon source}
As shown in Fig.~2 in the main text, high-quality entangled photon pairs are created from a polarization Sagnac interferometer (PSI) \cite{Kim06,Chaisson22} and used to support our theoretical predictions faithfully. The photon pairs generated from PSI, denoted as $\rho_{\text{expt}}$, are close to the target state
\begin{eqnarray}
\ket{\Psi^{-}}&=&\frac{1}{\sqrt{2}}(\ket{H}_{A}\ket{V}_{B}-\ket{V}_{A}\ket{H}_{B})\\
&=&\frac{1}{\sqrt{2}}(\ket{0}_{3}\ket{1}_{3}-\ket{1}_{3}\ket{0}_{3}).
\end{eqnarray}
We experimentally characterize $\rho_{\text{expt}}$ via state tomography using a pair of polarization analyzers, each consisting of a half-wave plate (HWP), a quarter-wave plate (QWP), and a polarizing beam splitter (PBS), in mode $A$ and mode $B$, respectively. The measured density matrix of the created photon pairs is:
\begin{widetext}
\begin{equation}
\rho_{\text{expt}}=
\left[
\begin{matrix}
0.021 & -0.012-0.008i & 0.013+0.011i & -0.002+0.002i\\
-0.012+0.008 & 0.541 & -0.488+0.026i & -0.001+0.001i\\
0.013-0.011i &  -0.488-0.026i  & 0.453 & 0.003+0.005i\\
-0.002-+0.002i & -0.001-0.001i & 0.003-0.005i & 0.004\\
\end{matrix}
\right].
\label{rho_90}
\end{equation}
\end{widetext}
The maximum-likelihood technique \cite{James01} has been used to determine the density matrix in our experiments. The fidelity of $\rho_{\text{expt}}$ and the target state, $\ket{\Psi^{-}}\!\!\bra{\Psi^{-}}$, is $F=\text{tr}\left(\rho_{\text{expt}}\ket{\Psi^{-}}\!\!\bra{\Psi^{-}} \right)\sim(98.62 \pm 0.26)\%$.

\section*{Supplementary Note 3: Figure 4 in the main text}

\subsection*{Fig. 4a}

In the Sagnac interferometer, as the two oppositely propagating photon pairs do not coherently interfere at $\text {PBS}$, the joint state of the photon in mode $A$ and the photon in mode $B$ will be in a completely mixed state:
\begin{equation}
\rho_\text{sep}=\frac{1}{2}(\ket{0}_{\!33}\!\!\bra{0}\!\otimes\!\ket{1}_{\!33}\!\!\bra{1}\!+\!\ket{1}_{\!33}\!\!\bra{1}\!\otimes\!\ket{0}_{\!33}\!\!\bra{0}).
\end{equation}
In Fig.~4\textbf{a} in the main text, we show how the experimental RSP capability of $\rho_\text{expt}$ is affected by $\rho_\text{sep}$. We therefore consider the following state as our theoretical state shared between Alice and Bob:
\begin{equation}
\rho(p_{\phi})=p_{\phi}\rho_\text{expt}+(1-p_{\phi})\rho_\text{sep},\label{rhophi}
\end{equation}
 where $0\leq p_{\phi}\leq 1$, as shown in Eq.~(18) in the main text.

In practical implementations, since $\rho_\text{sep}$ can be represented as a mixture of the states $\ket{\Psi^{-}}\!\bra{\Psi^{-}}$ and $\ket{\Psi^{+}}\!\bra{\Psi^{+}}$, i.e., $\rho_\text{sep}=(\ket{\Psi^{-}}\!\bra{\Psi^{-}}+\ket{\Psi^{+}}\!\bra{\Psi^{+}})/2$, where $\ket{\Psi^{+}}=(\ket{0}_3 \ket{1}_3+\ket{1}_3 \ket{0}_3)/\sqrt{2}$, we experimentally create the states for RSP with respect to the state $\rho(p_{\phi})$~(\ref{rhophi}) by setting different duration times for generating the $\ket{\Psi^{-}}\!\bra{\Psi^{-}}$ and $\ket{\Psi^{+}}\!\bra{\Psi^{+}}$. Let us take $p_{\phi}=0.8$ as a concrete example. Since the ratio of the states involving $\ket{\Psi^{-}}\!\bra{\Psi^{-}}$ (including $\rho_{{\rm expt}}$) to $\ket{\Psi^{+}}\!\bra{\Psi^{+}}$ in $\rho(p_{\phi})$ is
\begin{equation}
\frac{1}{2}(1+p_{\phi}):\frac{1}{2}(1-p_{\phi})=9:1,
\end{equation}
the ratio of the state creation duration times of $\ket{\Psi^{-}}\!\!\bra{\Psi^{-}}$ for RSP to the duration times of generating the $\ket{\Psi^{+}}\!\!\bra{\Psi^{+}}$ in our experiment is $9:1$ as well. They are respectively $18$ and $2$ seconds in our experiments.

\subsection*{Fig. 4b}

Figure~4\textbf{b} describes the experimental realizations of the RSPs using the Werner state:
\begin{equation}
\rho_{W}(p_{\text{noise}})=(1-p_{\text{noise}})\rho_{{\rm expt}}+\frac{p_{\text{noise}}}{4}I\otimes I,\label{wstate}
\end{equation}
for $0\leq p_{\text{noise}}\leq 1$. See Eq.~(19) in the main text. We use the same method in the cases for Fig.~4\textbf{a} to create the theoretical states $\rho_{W}(p_{\text{noise}})$. Since $I\otimes I/4$ in Eq.~(\ref{wstate}) can be decomposed into: $I\otimes I/4=(\ket{\Phi^{+}}\!\bra{\Phi^{+}}+\ket{\Phi^{-}}\!\bra{\Phi^{-}}+\ket{\Psi^{+}}\!\bra{\Psi^{-}}+\ket{\Psi^{-}}\!\bra{\Psi^{-}})/4$, the ration of the duration times for creating the states involving the $\ket{\Psi^{-}}\!\bra{\Psi^{-}}$ and the duration times for creating the other states in $\rho_{W}(p_{\text{noise}})$ is
\begin{equation}
(1-\frac{3}{4}p_{\text{noise}}):\frac{1}{4}p_{\text{noise}}.
\end{equation}
The actual duration times chosen in our experiments are therefore set accordingly for each case of $p_{\text{noise}}$.

\section*{Supplementary Note 4: Quantum discord and EPR steerability of the created photon pairs}

In this section, we utilize the geometric discord $\mathcal{D}$ \cite{Dakic10} and steerable weight $SW$ \cite{skrzypczyk2014} to measure quantum discord and EPR steerability of the created photon pairs, respectively. We will show how to calculate the geometric discord and steerable weight in the following subsections.

\subsection*{Geometric discord}

For a given $\rho_{\text{expt}}$, the geometric discord is calculated by the Bloch vector $\vec{x}\equiv\left( x_{1},x_{2},x_{3} \right) $ with $x_{i}=\text{Tr}\left[ \rho_{\text{expt}}\left(\sigma_{i}\otimes I \right)\right]$ and the correlation tensor $A$ with $A_{ij}=\text{Tr}\left[\rho_{\text{expt}}\left(\sigma_{i}\otimes \sigma_{j}\right)\right]$ \cite{Dakic10}. Here, $\{ \sigma_{i(j)} \}^{3}_{i(j)=1}$ is composed of $\sigma_{1}=X$, $\sigma_{2}=Y$, and $\sigma_{3}=Z$ Pauli matrices. The geometric discord of $\rho_{\text{expt}}$ takes the following form \cite{Dakic10}
\begin{equation}
\mathcal{D}(\rho_{\text{expt}})=\frac{1}{4}\left[ \text{Tr}\left(\vec{x}^{T}\vec{x}\right)+\text{Tr}\left( A^{T}A \right)-k_{\text{max}} \right],
\label{discord}
\end{equation}
where the superscript $T$ indicates the transpose operation and $k_{\text{max}}$ is the largest eigenvalue of matrix $K=\vec{x}\vec{x}^{T}+AA^{T}$. Thus, we can obtain the experimental geometric discord, denoted as $\mathcal{D}_{\text{expt}}$, by $\rho_{\text{expt}}$.

As the results shown in Fig.~4\textbf{a} in the main text, the density matrix of the created photon pairs at $p_{\phi}=0.4$ in our experiment is of the following form:
\begin{widetext}
\begin{equation}
\rho_{\text{expt}}(0.4)=
\left[
\begin{matrix}
    0.010 & -0.003 - 0.037i & -0.008 + 0.055i & -0.006 - 0.016i\\
  -0.003 + 0.037i &  0.470 & -0.193 + 0.006i &  0.011 - 0.050i\\
  -0.008 - 0.055i & -0.193 - 0.006i &  0.504 & 0.007 + 0.054i\\
  -0.006 + 0.016i &  0.011 + 0.050i &  0.007 - 0.054i  & 0.014
\end{matrix}
\right],
\label{rho_40}
\end{equation}
\end{widetext}
where the fidelity of $\rho_{\text{expt}}(0.4)$ and the target state, $\ket{\Psi^{-}}\!\bra{\Psi^{-}}$, is $F \sim 0.68$ in our experiment. The resulting geometric discord of $\rho_{\text{expt}}(0.4)$ is $\mathcal{D}_{\text{expt}}\sim 0.054$. Similarly, the theoretical geometric discord, $\mathcal{D}_{\text{theory}}\sim 0.075$, can be obtained by $\rho(0.4)$. Therefore, the state of the created photon pairs has quantum discord in terms of geometric discord; however, the resulting RSP cannot satisfy the fidelity criterion~(16) in the main text for $\bar{F}_{s,{\rm theory}}=78.62 \%$ and $\bar{F}_{s,{\rm expt}}=(78.27\pm0.03) \%$. With the same method as shown above, we obtain ($\mathcal{D}_{\text{theory}}\sim0.118$, $\mathcal{D}_{\text{expt}}\sim0.156$) of $\rho_{W}(p_{\text{noise}})$ for $p_{\text{noise}}=0.5$ compared to $\alpha=\beta=0$ shown in Fig.~4\textbf{b}. This demonstrates that a classical RSP can even be implemented with a $\rho_{W}(p_{\text{noise}})$ which possesses quantum discord.

\subsection*{Steerable weight}
Given Alice’s choice of the $m$th measurement and $n$th outcome, any assemblage $\{\rho'_{nm}\}_{nm}$ of unnormalized states which Alice's measurement affects Bob's particle into can be decomposed as
\begin{equation}
\rho'_{nm}=\mu\rho'_{nm,\text{US}}+(1-\mu)\rho'_{nm,\text{S}}  \ \ \ \ \forall n,m ~,
\end{equation}
where $0\leq\mu\leq1$ and $\{\rho'_{nm,\text{US}}\}_{nm}$ ($\{\rho'_{nm,\text{S}}\}_{nm}$) is the unsteerable (steerable) assemblage in which the elements can (cannot) be represented in the form of the local hidden state (LHS) model \cite{Wiseman07,Reid09}:
 \begin{equation}
\rho'_{nm,\text{US}}=\sum_{\lambda}p(v_{nm}|\lambda)\rho'_{\lambda} \ \ \ \ \ \ \ \forall n,m.\label{rp}
\end{equation}
The steerable weight $SW(\rho_{\text{expt}})$ is then defined as $SW(\rho_{\text{expt}})=1-\mu^*$, where $\mu^*$ denotes the maximum $\mu$ and can be obtained via SDP; see Ref.~\cite{skrzypczyk2014} for more detail. Here, to compare with the RSP capability that is measured by the process matrix of the RSP protocol constructed via the process tomography algorithm, we choose the measurements of the three Pauli matrices, $\sigma_1\ (X)$, $\sigma_2\ (Y)$, and $\sigma_3\ (Z)$ for $m=1,2,3$, respectively. Through the density matrices $\rho_{\text{expt}}(0.4)$ and $\rho(0.4)$, we can get the experimental and theoretical steerable weight, $SW_{\text{expt}}$ and $SW_{\text{theory}}$. As illustrated in Fig.~4\textbf{a} in the main text, the steerable weight of $\rho(0.4)$ and $\rho_{\text{expt}}(0.4)$ shown in Eq.~({\ref{rho_40}}) are respectively $SW_{\text{theory}}\sim0.183$ and $SW_{\text{expt}}\sim0.095$ which implies that the state of the created photon pairs has EPR steerability in terms of steerable weight, however, the resulting RSP cannot satisfy the fidelity criteria (16) in the main text.

\renewcommand{\refname}{Supplementary References}
